\documentclass[11pt,a4paper]{article}
\usepackage[utf8]{inputenc}\usepackage[T1]{fontenc} 
\usepackage{jheppub}
\usepackage{amsmath}
\usepackage{amssymb}
\usepackage{upgreek}
\renewcommand{\aa}{{\rm a}}
\usepackage{slashed}
\usepackage{graphicx}
\usepackage{subcaption}
\usepackage{eso-pic}
\usepackage{slashed}
\usepackage{cases}
\usepackage[toc,page]{appendix}
\usepackage{empheq}
\usepackage[normalem]{ulem} 
\usepackage{xcolor} 
\usepackage{afterpage}
\usepackage{float}
\usepackage{bbold}
\usepackage{hypbmsec}
\usepackage{tikz}
\usepackage[compat=1.1.0]{tikz-feynman}  
\usepackage{slashed}                     
\usepackage{changepage}
\usepackage{comment}
\usepackage{soul} 

\usepackage[export]{adjustbox}

\makeatletter
\newcommand\fake@math{}
\def\fake@math#1\){[math]}
\makeatother

\newcommand{\s}{\text{s}}
\newcommand{\sN}{\text{s}_{\N1}}
\newcommand{\sNN}{\text{s}_{\N2}}
\newcommand{\sn}{\text{s}_{\nu1}}
\newcommand{\snn}{\text{s}_{\nu2}}
\renewcommand{\c}{\text{c}}
\newcommand{\cN}{\text{c}_{\N1}}
\newcommand{\cNN}{\text{c}_{\N2}}
\newcommand{\cn}{\text{c}_{\nu1}}
\newcommand{\cnn}{\text{c}_{\nu2}}
\newcommand{\m}{m}
\newcommand{\U}{U}
\newcommand{\V}{\text{V}}
\newcommand{\Y}{\text{Y}}
\renewcommand{\O}{\text{O}}

\newcommand{\Uai}[2]{U_{#1#2}}
\newcommand{\Ua}[1]{U_{#1}}
\newcommand{\Ui}[1]{U_{#1}}
\newcommand{\Vai}[2]{\text{V}_{#1#2}}

\newcommand{\N}{\text{N}}
\newcommand{\Nsl}{\text{N}}
\newcommand{\Be}{e}

\renewcommand{\r}[2]{\text{r}_{#1}^{#2}}
\renewcommand{\Re}[1]{\operatorname{Re}(#1)}
\renewcommand{\Im}[1]{\operatorname{Im}(#1)}
\newcommand{\Mbar}{\bar{M}}

\newcommand{\labelo}[1]{ }

\allowdisplaybreaks

\begin{document}

\begingroup\raggedleft 
\hfil{IRMP-CP3-24-21} \\
\hfil{ZTF-EP-24-11} \\
\endgroup
\vskip 1.5cm

\title{On the collider-testability of the type-I seesaw model with 3 right-handed neutrinos}

\author[a]{Marco Drewes,}
\author[a]{Yannis Georis,}
\author[b,c,d,a]{Juraj Klari\'{c}}
\author[e]{and Antony Wendels}
\affiliation[a]{Centre for Cosmology, Particle Physics and Phenomenology (CP3), 
Universit\'{e} catholique de Louvain, Chemin du Cyclotron 2, 
B-1348 Louvain-la-Neuve, Belgium}
\affiliation[b]{Institute of Physics and Delta Institute for Theoretical Physics, University
of Amsterdam,\\Science Park 904, 1098 XH Amsterdam, The Netherlands}
\affiliation[c]{Theory Group, Nikhef,\\ Science Park 105, 1098 XG, Amsterdam, The Netherlands}
\affiliation[d]{Department of Physics, Faculty of Science, University of Zagreb,\\
Bijenicka c. 32, 10000 Zagreb, Croatia}
\affiliation[e]{Ecole Normale Supérieure Paris-Saclay,\\4 Av. des Sciences, 91190 Gif-sur-Yvette, France}

\emailAdd{marco.drewes@uclouvain.be}
\emailAdd{yannis.georis@uclouvain.be}
\emailAdd{juraj.klaric@nikhef.nl}
\emailAdd{antony.wendels@ens-paris-saclay.fr}

\abstract{If any heavy neutral leptons are discovered in accelerator-based experiments, key questions will involve their possible connection to neutrino masses or leptogenesis. Working in a renormalisable extension of the Standard Model by three right-handed neutrinos, we address the question of how much information about the fundamental model parameters can be obtained by measuring the branching ratios in the decays of the heavy neutral leptons into individual SM generations. We find that, provided that these branching ratios could be measured with arbitrary precision and assuming kinematically distinguishable right-handed neutrinos, they can be sufficient to pin down all 18 parameters of the model when supplemented with light neutrino oscillation data. When considering a finite statistical uncertainty comparable to that which can be achieved by future lepton colliders like FCC-ee or CEPC in the mass range of tens of GeV, some parameter degeneracies remain, but measurements would still provide powerful consistency checks of the model. In the sub-GeV range 
a good sensitivity to individual model parameters can be expected for SHiP and potentially for DUNE. This shows the potential of these experiments to not only discover heavy neutral leptons, but play an important role in understanding their role in particle physics and cosmology.
}

\maketitle

\section{Introduction}

Neutrino flavour oscillations to date comprise the sole established piece of evidence for  physics beyond the renormalisable Standard Model (SM) of particle physics which was  found in laboratory experiments.
A simple explanation for this observation is offered by extensions of the SM by right-handed neutrinos $\nu_R$. This idea is not only aesthetically and theoretically  motivated by noticing that all other known elementary fermions exist with both chiralities (and that those would be needed for anomaly-freedom in many gauge-extensions of the SM),
but also by the fact that the $\nu_R$ appear in many  neutrino mass models, in particular in the type-I seesaw mechanism \cite{Minkowski:1977sc, GellMann:1980vs, Mohapatra:1979ia, Yanagida:1980xy, Schechter:1980gr, Schechter:1981cv}.
In addition, they could potentially resolve several puzzles 
in particle physics and cosmology
that cannot be explained within the SM \cite{Boyarsky:2009ix,Drewes:2013gca,Abdullahi:2022jlv}, including the matter-antimatter asymmetry in the observable Universe \cite{Canetti:2012zc} through leptogenesis \cite{Fukugita:1986hr} or the existence of dark matter (DM)
\cite{Dodelson:1993je,Shi:1998km} (cf.~\cite{Davidson:2008bu,Bodeker:2020ghk,Klaric:2021cpi} and ~\cite{Drewes:2016upu,Boyarsky:2018tvu} respectively for overviews).

Once any heavy neutral leptons (HNLs) are discovered experimentally, the most burning questions will be to understand their role in particle physics and cosmology, 
in particular if and how they are connected to neutrino masses, baryogenesis, and potentially DM. 
The testability of this connection is model dependent and in particular limited by comparing the number of independent observables that can be extracted from data to the number of unknown model parameters. The latter strongly depends on the number $n_s$ of right-handed neutrino flavours. 
In the most general renormalisable Lagrangian
that can be constructed from $\nu_R$ and SM fields only (known as type-I seesaw Lagrangian) there are $7n_s - 3$ new parameters in addition to those in the SM.  
Since one $\nu_R$-flavour is needed for each non-zero light neutrino mass $m_i$ the minimal value consistent with experimental data is $n_s=2$ if one SM neutrino is massless and $n_s=3$ if all three of them are massive.\footnote{The $\nu$MSM in principle contains three flavours of $\nu_R$,  but observational constraints on the third HNL (which acts as a DM candidate) imply that baryogenesis and the seesaw mechanism are effectively described by $n_s=2$.} 
The minimal model with $n_s=2$ is known to be highly testable, and at last in principle all model parameters can be constrained by combining data from different experiments \cite{Hernandez:2016kel,Drewes:2016jae}.\footnote{Interestingly, this model can still lead to leptogenesis in the limit when the Majorana mass matrix of the $\nu_R$ is exactly degenerate \cite{Antusch:2017pkq}, leading to an even higher level of testability \cite{Sandner:2023tcg}.}

In the present work we address the question of how these conclusions change in the next-to-minimal scenario with $n_s=3$, still considering the most general renormalisable Lagrangian that can be constructed from the $\nu_R$ and SM-fields alone, to be defined in \eqref{L}.
We remain entirely agnostic regarding the flavour structure (or texture) of the $\nu_R$ mass and coupling matrices.\footnote{The dimension of the parameter space can be reduced by discrete symmetries \cite{King:2017guk,Xing:2020ijf,Chauhan:2022gkz,Chauhan:2023faf}, cf.~\textit{e.g.}\cite{Hagedorn:2014wha,Chauhan:2021xus,Drewes:2022kap,Drewes:2024pad} for a specific realisation.}
The question of testability can be addressed at several levels,
\begin{enumerate}
    \item\label{it:inPrinciple} Does a given set of observables \emph{in principle} contain enough information to reconstruct a given set of model parameters (and potentially overconstrain the model) in an idealised world where they can be measured with arbitrary precision? 
    \item\label{it:EventNumbers} Can any experiment in foreseeable future observe enough events such that the statistical uncertainties permit extracting this information, assuming idealised detectors and no systematic uncertainties?
    \item\label{it:Detectors} How much information can be extracted with realistic detectors?
\end{enumerate}
In the present work we address the first two of these questions, which can be done without realistic detector simulations. Further, we entirely restrict ourselves to counting numbers of events, more specifically, the branching ratios of the decay of each HNL $N_i$ into the SM generations $\alpha = e, \mu, \tau$. 
With $n_s=3$ flavours of HNLs, this leaves us with nine observables plus the three HNL masses $M_i$
to constrain eighteen model parameters. 
However, many of these can be determined or constrained from other sources. 
The light neutrino mass splittings and mixings have already been measured (five parameters), and also the absolute mass scale can in principle determined in low energy experiments \cite{Katrin:2024tvg} and cosmology \cite{Planck:2018vyg}. The Dirac phase is expected to be measured in the foreseeable future by DUNE \cite{DUNE:2015lol} or T2HyperK \cite{Hyper-Kamiokande:2022smq}, and a neutrinoless double beta decay is sensitive to a combination of the Majorana phases (and potentially phases in the HNL sector \cite{Bezrukov:2005mx,Asaka:2011pb,Lopez-Pavon:2012yda,Drewes:2016lqo,Hernandez:2016kel,deVries:2024rfh}). Hence, even when only counting the number of HNL decays and their branching ratios at colliders, this can be sufficient to formulate testable predictions.

This paper is organised as follows. We first review in Sec.~\ref{sec:modelandparam} the type-I seesaw Lagrangian and set up the notations. 
We also derive the fundamental domains for the model parameters. 
Then, in Sec.~\ref{sec:inprinciple_learning}, we address the question of what can in principle be learned for number of events at colliders in the limit where the branching ratios to the different SM flavours are known to infinite precision. 
In Sec.\ref{sec:chi2method}, we go beyond this idealised scenario and, using a standard $\chi^2$ approach, we derive semi-realistic limits on how much one could constrain model parameters if HNLs are discovered at the FCC-ee. 
We finally conclude in Sec.~\ref{sec:conclusion}. 
One appendix with analytic formula for the HNL flavour branching ratios is also provided.

\section{Model and parametrisation}
\label{sec:modelandparam}

In this section we specify the model and set our conventions. We  define a particularly useful incarnation of the Casas-Ibarra parametrisation \cite{Casas:2001sr} that was introduced in \cite{Drewes:2021nqr}, cf.~\eqref{eq:OROparametrisation} and \eqref{OROParametrisation} below. 

\subsection{Review of basic ingredients}
\label{subsec:reviewingredients}

When it comes to the experimental testability of seesaw models, one important question is whether the new HNLs $N_i$ can be found experimentally. 
This first and foremost depends on whether the seesaw scale $\Mbar$ associated with their masses $M_i$ is kinematically accessible in accelerator-based experiments. 
While $\Mbar$ it is traditionally associated with  scales far beyond collider reach (\textit{e.g.}~related to of Grand Unification), 
technically natural models with $\Mbar$ at and below the TeV scale exist (cf.~\textit{e.g.}~Sec.~5 in \cite{Agrawal:2021dbo} and references therein). 
Different searches have been proposed (cf.~\textit{e.g.}~\cite{Gorbunov:2007ak,Atre:2009rg,Drewes:2013gca,Deppisch:2015qwa,Antusch:2016ejd,Abada:2017jjx,Chrzaszcz:2019inj,Abdullahi:2022jlv} and references therein), several of which have been and are being performed at various facilities (cf.~\cite{Abdullahi:2022jlv,Antel:2023hkf} and references therein). Amongst the more recent results are \textit{e.g.}~those obtained by ATLAS \cite{ATLAS:2022atq}, CMS \cite{CMS:2023jqi,CMS:2024ita,CMS:2024xdq,CMS:2024ake,CMS:2024hik}, T2K \cite{T2K:2019jwa}, Belle \cite{Belle:2024wyk}, and NA62 \cite{NA62:2020mcv}.
In the current work we assume that the $N_i$ can be produced at colliders, but no other new particle that they couple to is accessible. 
We furthermore restrict ourselves to renormalisable interactions as described by the Lagrangian \eqref{PhenoModelLagrandian} below, which may in principle be all there is up to the Planck scale $m_P$ \cite{Shaposhnikov:2007nj,Bezrukov:2014ina}, but on a more general basis can  be viewed as the lowest order of an effective field theory expansion \cite{delAguila:2008ir,Liao:2016qyd,DeVries:2020jbs}. The neutrino minimal Standard Model ($\nu$MSM) \cite{Asaka:2005an,Asaka:2005pn} represents an example in which neutrino oscillations, leptogenesis and DM can all be resolved for $\Mbar$ below the electroweak scale with no other new particles up to $m_P$ \cite{Canetti:2012kh,Ghiglieri:2020ulj}.

The second important question is whether the production cross section for $N_i$ is sufficient to produce them in sizeable numbers.
In the minimal scenario, the $\nu_R$ interact with ordinary matter only through their mixing with the SM neutrinos, which is also responsible for the generation of the light neutrino masses. 
More precisely, the mass eigenstates $N_i \simeq \nu_{R i} + \Theta_{\alpha i} \nu_{L \alpha}^c$
couple to the weak currents with a strength that is suppressed by the elements  of the active-sterile mixing matrix $\Theta_{\alpha i}$ \cite{Shrock:1980ct,Atre:2009rg},
\begin{equation}
 \mathcal L
\supset
- \frac{m_W}{v} \overline N_i \Theta^*_{\alpha i} \gamma^\mu e_{L \alpha} W^+_\mu
- \frac{m_Z}{\sqrt 2 v} \overline N_i \Theta^*_{\alpha i} \gamma^\mu \nu_{L \alpha} Z_\mu
- \frac{M_i}{v} \Theta_{\alpha i} h \overline{\nu_L}_\alpha N_i
+ \text{h.c.}
\ ,\label{PhenoModelLagrandian}
\end{equation}
with $v$ the Higgs field expectation value and $m_Z$, $m_W$ the weak gauge boson masses.

For the proof-of-principle in the present work,
we assume that all three HNLs can be distinguished kinematically in experiments. In this regime the HNL production and decay can be treated as independent processes, in particular implying that we do not consider effects that come from interferences between channels with different $N_i$-flavours when simply counting the number of events in each decay channel.\footnote{In principle, more complicated observables can be used, such as angular and energy distributions, lepton number violation, or correlations between various observables. 
While such processes in principle provide an attractive target for collider searches (cf.~\textit{e.g.}~\cite{Anamiati:2017rxw,Das:2017hmg,Dib:2017iva,Cvetic:2018elt,Hernandez:2018cgc,Abada:2019bac,Drewes:2019byd,Abada:2022wvh,Antusch:2023jsa}), their simulation requires non-standard tools \cite{Antusch:2022ceb} and goes beyond the scope of this work.} 
Under this assumption we can factorise the HNL production and decay, with their production cross section $\sigma_{N_i}$ and their total decay width $\Gamma_{N_i}$ both controlled by the magnitudes of the mixing angles,
\begin{eqnarray}\label{UaiDefinition}
U_{\alpha i}^2 \equiv |\Theta_{\alpha i}|^2,
\end{eqnarray}
and approximately given by \cite{Atre:2009rg}
\begin{eqnarray}\label{eq:sigmaN}
\sigma_{N_i} =  \sum_{\alpha=e,\mu,\tau} U_{\alpha i}^2\sigma_{\nu_\alpha}\Pi \ , \quad
\Gamma_{N_i} \simeq \sum_{\alpha=e,\mu,\tau}  \frac{U_{\alpha i}^2}{96\pi^3}   M^5 G_F^2\aa.
\end{eqnarray}
Here $\sigma_{\nu_\alpha}$ is the production cross section for SM neutrinos $\nu_{L\,\alpha}$, $\Pi$ is a kinematic factor that takes into account the  mass of the HNLs, $G_F$ is the standard Fermi constant, $\aa\simeq 12$, and all final state masses are neglected. A list of Feynman processes for the HNL production and decay at colliders can be found in \textit{e.g.}~\cite{Antusch:2016ejd}.
As far as the questions \ref{it:inPrinciple} and \ref{it:EventNumbers} are concerned we can, under the aforementioned assumptions, assume that measuring HNL lifetimes and branching ratios can directly be translated into a determination of the $U_{\alpha i}^2$ (and vice versa). By focusing on the $U_{\alpha i}^2$ alone we can perform the study semi-analytically at a proof-of-principle level.

One may naively expect that the $U_{\alpha i}^2$ are suppressed by the ratio between the light neutrino and HNL masses,
\begin{eqnarray} \label{NaiveSeesaw}
    U_0^2 \equiv \frac{\sum_i m_i}{\Mbar} \ 
    \ll 1.
\end{eqnarray}
However, natural models can yield mixing angles of order one if the light neutrino masses are protected by a symmetry \cite{Shaposhnikov:2006nn,Kersten:2007vk,Moffat:2017feq}.
We shall quantify this enhancement by a small parameter $\upepsilon$ 
\begin{eqnarray}\label{eq:enhancement}
U_{\alpha i}^2/U_0^2  \sim 1/\upepsilon,
\end{eqnarray}
the precise definition we provide below in \eqref{UpepsilonDef}.
Examples of such symmetry-protected scenarios include the inverse seesaw \cite{Mohapatra:1986aw,Mohapatra:1986bd,Bernabeu:1987gr}, linear seesaw \cite{Akhmedov:1995ip,Akhmedov:1995vm} and the $\nu$MSM \cite{Asaka:2005an,Asaka:2005pn}. 
It turns out that the 
range of masses and mixings for which the $\nu_R$ can simultaneously explain the neutrino masses and the observed matter-antimatter asymmetry 
is indeed accessible by accelerator-based experiments \cite{Canetti:2012kh,Klaric:2021cpi,Hernandez:2022ivz,Drewes:2021nqr}.

In order to connect the $M_i$ and $U_{\alpha i}^2$ to data from neutrino oscillation experiments we consider the seesaw Lagrangian
\begin{eqnarray}
	\label{L}
	\mathcal L \supset
	\mathrm{i} \, \overline{\nu_R} \, \slashed\partial \, \nu_{R}
	- \frac{1}{2}
	\overline{\nu^c_R}\, M_M\, \nu_{R}
	- \overline{l_{L}} \, Y \, \varepsilon \Phi^* \, \nu_{R}
	+ {\rm h. c.} \, .
\end{eqnarray}
with $M_M$ the Majorana mass matrix of the $\nu_R$, $Y$ the neutrino Yukawa coupling matrix, $\Phi$ the SM Higgs doublet, $\varepsilon$ the totally antisymmetric SU(2) tensor and $l_L=( \nu_L,  e_L )^T$  the three LH lepton doublets (with the lepton flavour index $\alpha=e,\, \mu,\, \tau$). 
Throughout this paper, we shall assume that the HNLs are kinematically distinguishable at colliders. In this regime, the correction to the HNL masses from the Higgs mechanism (which is parametrically of the same order as the light neutrino masses) can be neglected. 
In this context, the $M_i$ are in good approximation given by the eigenvalues of $M_M$, and $\Theta \simeq \theta \equiv v Y^\dagger M_M^{-1} $.

\subsection{A convenient parametrisation}

The connection between the Lagrangian \eqref{L} and its low energy limit \eqref{PhenoModelLagrandian} to the light neutrino properties can be made by the Casas-Ibarra parametrisation \cite{Casas:2001sr},
\begin{equation}\label{CAparam}
    \Theta \simeq 
    i V_{\nu} \sqrt{m_\nu^{\mathrm{diag}}} \mathcal{R} \sqrt{\rm M_M}^{-1}  ,
\end{equation}
where $\mathcal{R}$ is a complex $3\times n_s$ matrix verifying $\mathcal{R} \mathcal{R}^T = \mathbb{1}$, $m_\nu^{\mathrm{diag}} = \mathrm{diag}(m_i)$ represents the light neutrino mass matrix in their mass basis, and $V_\nu$ is the light neutrino mixing matrix, or PMNS matrix, that diagonalises $m_\nu \approx -\theta M_M^{-1}\theta^T$, which we parameterise as
\begin{eqnarray}
\label{PMNS}
V_\nu=V^{(23)}U_\delta V^{(13)}U_{-\delta}V^{(12)}{\rm diag}(e^{ i \alpha_1/2},e^{i \alpha_2 /2},1)\,,
\end{eqnarray}
with $U_{\pm \delta}={\rm diag}(e^{\mp i \delta/2},1,e^{\pm i \delta /2})$.
The non-zero entries of the matrices $V^{(ij)}$ are given by
\begin{eqnarray}
V^{(ab)}_{aa}=V^{(ab)}_{bb}=\cos \uptheta_{ab} \ , \
V^{(ab)}_{ab}=-V^{(ab)}_{ba}=\sin \uptheta_{ab} \ , \
V^{(ab)}_{cc}=1 \quad \text{for $c\neq a,b$},
\end{eqnarray}
with $\uptheta_{ab}$ the light neutrino mixing angles.

A convenient way of parametrising $\mathcal{R}$ is the following for $n_s=3$ \cite{Drewes:2021nqr},
\begin{equation}
    \mathcal{R} = \O_\nu \mathrm{R_C} \O_\N  ,
    \label{eq:OROparametrisation}
\end{equation}
where  $\O_\nu$ and $\O_\N$ are SO$_3$($\mathbb{R}$) matrices and $\mathrm{R_C}$ is a SO$_3$($\mathbb{C}$) matrix. Here we introduce four real angles $\theta_{\nu 1},~\theta_{\nu 2},~\theta_{\N 1},~\theta_{\N 2}$ and a complex angle $\omega + i\gamma$ such that
\begin{equation}\label{OROParametrisation}
\begin{aligned}
    \O_\nu =
    \begin{pmatrix}
        \cnn & 0 & \snn \\
        0 & 1 & 0 \\
        - \snn & 0 & \cnn
    \end{pmatrix} \cdot&
    \begin{pmatrix}
        1 & 0 & 0 \\
        0 & \cn & \sn \\
        0 & -\sn & \cn
    \end{pmatrix} , \hspace{.5cm} \O_\N =
    \begin{pmatrix}
        1 & 0 & 0 \\
        0 & \cN & \sN \\
        0 & -\sN & \cN
    \end{pmatrix}\cdot 
    \begin{pmatrix}
        \cNN & 0 & \sNN \\
        0 & 1 & 0 \\
        - \sNN  & 0 & \cNN 
    \end{pmatrix} 
    \\
    &\mbox{ and } \mathrm{R_C} =
    \begin{pmatrix}
        \cos{(\omega + i \gamma)} & \sin{(\omega + i \gamma)} & 0 \\
        - \sin{(\omega + i \gamma)} & \cos{(\omega + i \gamma)} & 0 \\
        0 & 0 & 1
    \end{pmatrix}.
\end{aligned}
\end{equation}
The quantities $\c_{\nu 1,2}$, $\s_{\nu 1,2}$, $\c_{\N 1,2}$ et $\s_{\N 1,2}$ represent cosine and sine of the angles $\theta_{\nu 1},~\theta_{\nu 2},~\theta_{\N 1}$ and $\theta_{\N 2}$ respectively. 
An important property of the parametrisation \eqref{OROParametrisation} is that the overall size \eqref{UpepsilonDef} of the active-sterile mixing $U_{\alpha i}^2/U_0^2$ relative to \eqref{NaiveSeesaw} can be expressed in terms of the imaginary part of one single parameter in $\mathrm{R_C}$  alone,\footnote{Furthermore, this way of parametrising $\mathcal{R}$ resembles the one of the scenario $n_s=2$; for instance, $\mathcal{R}$ corresponds only to the first two columns of $\mathrm{R_C}$ for normal ordering.}
\begin{eqnarray}\label{UpepsilonDef}
\upepsilon = e^{-2\gamma} 
\end{eqnarray}

In the model with $n_s=2$ the $U_{\alpha i}^2$ are polynomials in $\upepsilon$ only involving integer powers $\upepsilon^1$, $\upepsilon^0$ and  $\upepsilon^{-1}$ \cite{Asaka:2011pb,Ruchayskiy:2011aa,Drewes:2016jae}. 
A discovery of the HNLs is possible because of the $1/\upepsilon$-terms, and some of their properties are accessible at this order, such as the ratios $U_\alpha^2/U^2$ with
\begin{eqnarray}
\label{U2defs}
\ U_i^2 = \sum_\alpha U_{\alpha i}^2 \quad , \quad U_\alpha^2 = \sum_i U_{\alpha i}^2 \quad , \quad U^2 = \sum_i U_i^2 = \sum_\alpha U_\alpha^2 \; .
\end{eqnarray}
It is further convenient to define the analogue quantities
\begin{equation}\label{UaTildeDef}
    \Mbar \tilde{\U}_{\alpha}^2 \equiv
    \sum_{i} M_i \Uai{\alpha}{i}^2 = 
    [\V_{\nu} \sqrt{m_\nu^{diag}} \O_\nu \mathrm{R_C} \mathrm{R_C}^\dagger \O_\nu^\dagger \sqrt{m_\nu^{diag}} \V_\nu^\dagger]_{\alpha \alpha} ,
\end{equation}
as well as
\begin{equation}\label{UiTildeDef}
    \Mbar \tilde{\U}_{i}^2 \equiv \sum_{\alpha} M_i \Uai{\alpha}{i}^2 = [\O_\N^\dagger \mathrm{R_C}^\dagger \O_\nu^\dagger m_\nu^{diag} \O_\nu \mathrm{R_C} \O_\N]_{i i}
\end{equation}
and
\begin{equation}\label{eq:MU2Def}
    \tilde{\U}^2 =  \sum_{\alpha}\tilde{\U}_\alpha^2
    =  \sum_i\tilde{\U}_i^2.
\end{equation}
The combinations 
\eqref{UaTildeDef}  
and
\eqref{UiTildeDef}
are independent of the HNL masses $M_i$, and the $\tilde{U}_i^2$ as well as $\tilde{U}_\alpha^2$ only depend on the overall scale $\Mbar \equiv (M_1 + M_2 + M_3)/3$.
Further, the  $\tilde{\U}_{\alpha}^2$ are independent of $\theta_{\N 1,2}$, and the $\tilde{\U}_{i}^2$ are independent of the phases and angles in $V_\nu$.
Hence, the rest of our analysis should remain valid for any HNL masses (and mass splittings) as long as the HNLs can be kinematically resolved and the number of observed events is fixed. In the limit where all three HNLs are degenerate, the $\tilde{U}_{\alpha}^2,\tilde{U}_i^2$ coincide with the usual $U_\alpha^2$, $U_i^2$ defined in \eqref{U2defs}.

\subsection{Fundamental domain of the model parameters}

The parametrisation \eqref{CAparam} with \eqref{OROParametrisation} is redundant in the sense that there is more than one choice of the parameters that lead to exactly the same physics, \textit{i.e.}, the same  $Y$ and $M_M$ in \eqref{L}. Each line in Tab.~\ref{tab:param_restriction1} shows a choice of fundamental domains of the parameters that is sufficient to cover all possible values of $Y$ and $M_M$ (working in the basis where $M_M$ is diagonal). 
All parameter choices can be mapped onto any of these by a transformation that leaves $Y$ invariant. 
Each line in Tab.~\ref{tab:param_restriction2} 
corresponds to a choice of fundamental parameter domains that is sufficient to cover all possible values in $Y$ up to a discrete transformation, as indicated in the leftmost column. The exact sets of transformations leading to a maximal reduction of the parameters domain are specified in Tab.~\ref{tab:transfo_reduction}. These discrete transformations practically imply overall sign changes in the fields $\nu_{R i}$ or re-labelling\footnote{This in practice implies that one cannot fix the ordering of the HNL masses (\textit{e.g.} fixing $M_1<M_2<M_3$) but all possible orderings should be considered when doing a parameter space scan.} of the HNLs. Both of these operations are unphysical in the model \eqref{L}. However, this is in general not true if the $\nu_R$ possess additional interactions. Since we work in the framework of \eqref{L},  any of the choices of fundamental domains in Tab.~\ref{tab:param_restriction2} cover the entire physical parameter space of the model. In scenarios with new HNL interactions one may, depending on the shape of those operators, have to resort to Tab.~\ref{tab:param_restriction1} and a subset of the lines in Tab.~\ref{tab:param_restriction2} that still leaves the action invariant when those are included.

\begin{table}[!t]
    \centering
    \begin{tabular}{|ccccccc|}
        \hline
        $\theta_{\nu 1}$ & $\theta_{\nu 2}$ & $\theta_{\N 1}$ & $\theta_{\N 2}$ & $\omega$ & $\gamma$ & $\alpha_{1,2}$
        \\
        \hline \hline
    
        $[0,\pi]$, & $[0,\pi]$, & $[0,2\pi]$, & $[0,2\pi]$, & $[0,\pi]$, &$\mathbb{R}_+$, & $[0,4\pi]$
        \\
        $[0,\pi]$, & $[0,\pi]$, & $[0,\pi/2]$, & $[0,2\pi]$, & $[0,2\pi]$, &$\mathbb{R}$, & $[0,4\pi]$
        \\
        $[0,\pi]$, & $[0,\pi]$, & $[0,2\pi]$, &$[0,2\pi]$, &$[0,\pi/2]$, &$\mathbb{R}$, &$[0,4\pi]$
        \\
         $[0,\pi]$, & $[0,\pi]$, &$[0,\pi]$, &$[0,\pi]$, &$[0,2\pi]$, &$\mathbb{R}$, &$[0,4\pi]$
        \\
        $[0,\pi]$, & $[0,\pi]$, & $[0,\pi]$, &$[0,2\pi]$, &$[0,\pi]$, &$\mathbb{R}$, &$[0,4\pi]$
        \\
        \hline

    \end{tabular}
    \caption{Possible choices for the fundamental domains of the parameters in \eqref{CAparam} with \eqref{OROParametrisation} that covers all values of $Y$.}
    \label{tab:param_restriction1}
\end{table}

\begin{table}[!t]
    \centering
    \begin{tabular}{|c||ccccccc|}
        \hline
        Transformations & $\theta_{\nu 1}$ & $\theta_{\nu 2}$ & $\theta_{\N 1}$ & $\theta_{\N 2}$ & $\omega$ & $\gamma$ & $\alpha_{1,2}$
        \\
        \hline \hline
        
        $\Y \rightarrow \pm \Y$
        & $[0\,, \pi]$, &$[0\,, \pi]$, & $[0\,, 2\pi]$, &$[0\,, \pi]$, &$[0\,, \pi]$, &$\mathbb{R}_+$, &$[0\,, 4\pi]$
        \\
        & $[0\,, \pi/2]$, & $[0\,, \pi/2]$, &$[0\,, 2\pi]$, &$[0\,, 2\pi]$, &$[0\,, \pi]$, &$\mathbb{R}$, &$[0\,, 4\pi]$
        \\
        & $[0\,, \pi]$, & $[0\,, \pi]$, &$[0\,, \pi/2]$, &$[0\,, 2\pi]$, &$[0\,, \pi]$, &$\mathbb{R}$, &$[0\,, 4\pi]$
        \\
        & $[0\,, \pi]$, & $[0\,, \pi]$, &$[0\,, \pi]$, &$[0\,, \pi]$, &$[0\,, \pi]$, &$\mathbb{R}$, &$[0\,, 4\pi]$
        \\
        \hline
        
        $\Y \rightarrow \pm \Y\cdot \mathrm{P}$
        & $[0\,, \pi]$, &$[0\,, \pi]$, &$[0\,, \pi/2]$, &$[0\,, 2\pi]$, &$[0\,, \pi]$, &$\mathbb{R}_+$, &$[0\,, 4\pi]$
        \\
        & $[0\,, \pi/2]$, &$[0\,, \pi]$, &$[0\,, \pi]$, &$[0\,, 2\pi]$, &$[0\,, \pi]$, &$\mathbb{R}_+$, &$[0\,, 4\pi]$
        \\
        & $[0\,, \pi/2]$, &$[0\,, \pi/2]$, &$[0\,, \pi/2]$, &$[0\,, 2\pi]$, &$[0\,, 2\pi]$, &$\mathbb{R}$, &$[0\,, 4\pi]$
        \\
        \hline
        $\Y \rightarrow \Y\cdot \mathrm{P} \cdot$
        & $[0\,, \pi/2]$, &$[0\,, \pi]$, &$[0\,, \pi]$, &$[0\,, \pi/2]$, &$[0\,, \pi]$, &$\mathbb{R}_+$, &$[0\,, 4\pi]$
        \\
        $\cdot$diag($\pm 1$,$\pm 1$,$\pm 1$)
        & $[0\,, \pi/2]$, &$[0\,, \pi/2]$, &$[0\,, \pi/2]$, &$[0\,, \pi/2]$, &$[0\,, 2\pi]$, &$\mathbb{R}$, &$[0\,, 4\pi]$
        \\ \hline
    \end{tabular}
    \caption{Possible choices of the fundamental domains of parameters in  \eqref{CAparam} with \eqref{OROParametrisation} that cover the possible values of $Y$ up to transformations indicated in the leftmost column. }
    \label{tab:param_restriction2}
\end{table}

\begin{table}[!t]
\centering
\begin{tabular}{@{}c@{~}||c@{\hspace{.05cm}}c@{\hspace{.05cm}}c@{\hspace{.05cm}}c@{\hspace{.05cm}}c@{\hspace{.05cm}}c@{\hspace{.05cm}}c@{\hspace{.05cm}}c@{\hspace{.08cm}}||c@{\hspace{.05cm}}c@{\hspace{.05cm}}c@{}}
    Variables &$\theta_{\nu 1}$ &$\theta_{\nu 2}$& $ \omega$ & $\gamma$ & $\theta_{\N 1}$&$\theta_{N2}$ & $\alpha_1$&$\alpha_2$
    &$ \N_1$&$\N_2$&$\N_3$
    \\ 
    \hline
    Sent to &$2\pi - \theta_{\nu 1}$&$\theta_{\nu 2}$& $\pi - \omega$ & $-\gamma$ & $\pi + \theta_{\N 1}$ & $\theta_{N2}$&$\alpha_1$&$2\pi + \alpha_2$
    &$ -\N_1$& $-\N_2$&$-\N_3$
    \\
    &$\pi+\theta_{\nu 1}$ &$\pi - \theta_{\nu 2}$ &$ \omega$ &$\gamma$&$\theta_{\N 1}$& $\pi + \theta_{\N 2}$ & $2\pi + \alpha_1$ & $2\pi + \alpha_2$
    &$ -\N_1$&$\N_2$&$-\N_3$
    \\
    &$\pi - \theta_{\nu 1}$ &$\theta_{\nu 2}$&$ \omega$ &$\gamma$&$\pi- \theta_{\N 1}$ &$\theta_{N2}$&$\alpha_1$&$2\pi+\alpha_2$
    &$ \N_1$&$-\N_2$&$\N_3$
    \\
    &$\theta_{\nu 1}$& $\pi + \theta_{\nu 2}$ &$ \omega$ &$\gamma$&$\pi+\theta_{\N 1}$ &$\pi\pm\theta_{\N 2}$ &$\alpha_1$&2$\pi+\alpha_2$
    &$ \N_1$&$\N_2$&$-\N_3$
    \\
    &$\theta_{\nu 1}$&$\pi-\theta_{\nu 2}$ &$\pi-\omega$ &$-\gamma$& $2\pi- \theta_{\N 1}$ &$\theta_{N2}$&$2\pi+\alpha_1$ &$2\pi+\alpha_2$
    &$ -\N_1$&$\N_2$&$-\N_3$
    \\
    &$\theta_{\nu 1}$&$\theta_{\nu 2}$&$\pi+\omega$ &$\gamma$& $2\pi- \theta_{\N 1}$ &$2\pi-\theta_{\N 2}$&$\alpha_1$&$\alpha_2$
    &$ -\N_1$&$-\N_2$&$\pm\N_3$
    \\
    &$\theta_{\nu 1}$&$\theta_{\nu 2}$&$\pi+\omega$&$\gamma$&$\pi-\theta_{\N 1}$&$\theta_{N2}$&$\alpha_1$&$\alpha_2$
    &$ -\N_1$&$\N_2$&$-\N_3$
    \\
    &$\theta_{\nu 1}$&$\theta_{\nu 2}$&$ \omega$&$\gamma$&$\theta_{\N 1}$&$\pi+\theta_{\N 2}$&$\alpha_1$&$\alpha_2$
    &$ -\N_1$&$\N_2$&$-\N_3$
    \\
    &$\theta_{\nu 1}$&$\theta_{\nu 2}$&$ \omega$ &$\gamma$&$\theta_{\N 1}$&$-\frac{\pi}{2} + \theta_{\N 2}$&$\alpha_1$&$\alpha_2$
    &$ -\N_3$&$\N_2$&$\N_1$
\end{tabular}
    \caption{Set of transformations used to reduce the fundamental domain to the one shown in the last line of Tab.~\ref{tab:param_restriction2}.} 
    \label{tab:transfo_reduction}
\end{table}

\section{What can in principle be learnt from HNL lifetimes and branching ratios?}
\label{sec:inprinciple_learning}

We shall first address question \ref{it:inPrinciple}, i.e, how many of the $7n_s - 3$ model parameters in \eqref{L} can \emph{in principle} be extracted from
measuring the masses, lifetimes and branching ratios of HNLs, provided that all light neutrino parameters with the exception of the Majorana phases have been measured ?
For $n_s=2$ it is known that all 11 parameters can in theory be constrained from measurements of the $U_{\alpha i}^2$ at sufficiently high precision \cite{Hernandez:2016kel,Drewes:2016jae}, but this is challenging in practice because it requires an experiment that is sensitive to terms $\mathcal{O}[\upepsilon^0]$, \textit{i.e.}, can reach the seesaw line \eqref{NaiveSeesaw} in the HNL mass-mixing plane.
We now investigate the case $n_s=3$. 
With the aforementioned assumptions, the (eleven) unknown parameters comprise the three HNL masses $M_i$, the two Majorana phases $\alpha_1$ and $\alpha_2$ in $V_\nu$
and the four real angles $\theta_{\nu 1},~\theta_{\nu 2},~\theta_{\N 1},~\theta_{\N 2}$ plus one complex angle $\omega + i\gamma$ in \eqref{OROParametrisation}.

\begin{figure}[!t]
    \centering
    \includegraphics[width=.49\textwidth]{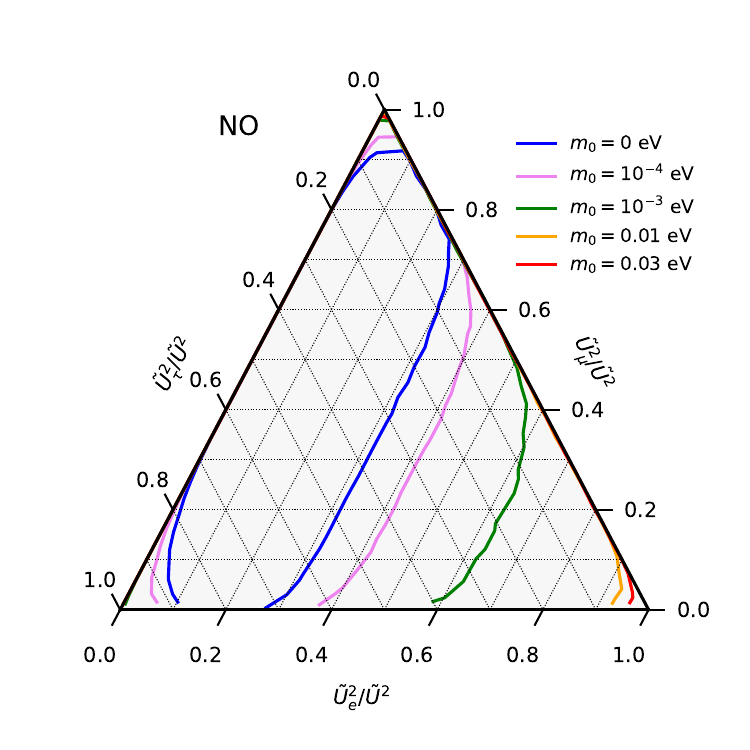}
    \includegraphics[width=.49\textwidth]{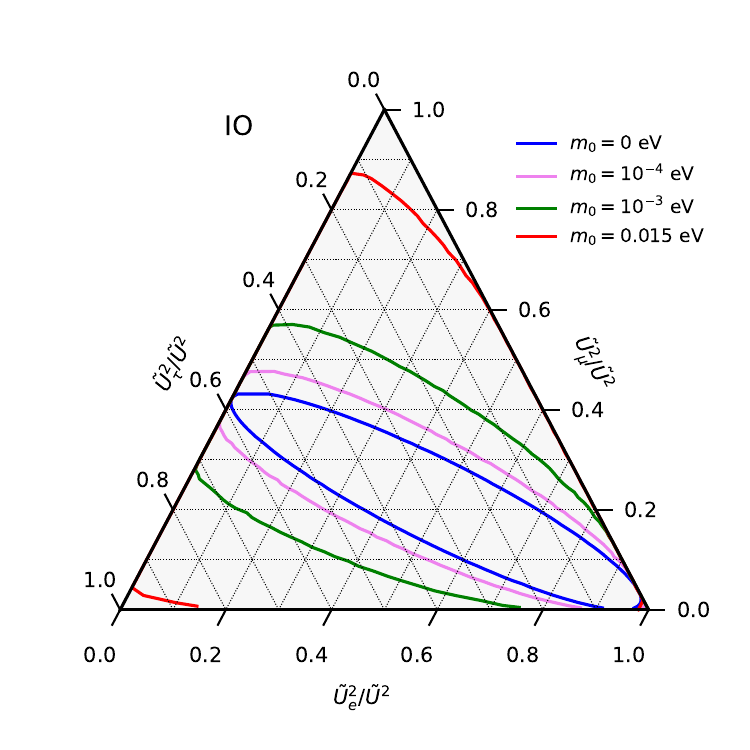}
    \caption{Range of flavour ratio $\tilde{U}_\alpha^2/\tilde{U}^2$ consistent with neutrino oscillation data for both normal (left plot) and inverted ordering (right plot) for different values of the lightest neutrino mass $m_0$. The three PMNS angles and 2 observed neutrino mass splittings are fixed to their best fit values \cite{Esteban:2020cvm} but $\delta$ is completely marginalised over. Compare with Fig.~\ref{fig:ternaryvanishingm0} to observe the impact of marginalising over the light neutrino parameters. 
    }
    \label{fig:ternarygeneral_multiplem0}
\end{figure}

A set of quantities that is know to be very useful are the flavour ratios 
\begin{equation}\label{Uratios}
     \hspace{-1cm}\frac{\tilde{U}_{\alpha}^2}{\tilde{U}^2} = \frac{\left[\V_{\nu} \sqrt{m_\nu^{diag}} \O_\nu \mathrm{R_C} \mathrm{R_C}^\dagger \O_\nu^\dagger \sqrt{m_\nu^{diag}} \V_\nu^\dagger\right]_{\alpha \alpha}}{\operatorname{Tr}\left[\V_{\nu} \sqrt{m_\nu^{diag}} \O_\nu \mathrm{R_C} \mathrm{R_C}^\dagger \O_\nu^\dagger \sqrt{m_\nu^{diag}} \V_\nu^\dagger\right]}.
\end{equation}
In the model with $n_s=2$ these ratios are determined at leading order by the light neutrino properties alone, providing not only a powerful consistency check, but also an indirect way of measuring the Majorana phase in the light neutrino mixing matrix $V_\nu$ \cite{Drewes:2016jae,Caputo:2016ojx}. 
Other parameters, such as the analogue to the real part $\omega$, become only accessible at order $\upepsilon^0$ for $n_s=2$, implying that determining them from measurements of the $U_{\alpha i}^2$ alone requires an experiment the  sensitivity of which can reach the so-called seesaw line defined by $U_0^2$ in \eqref{NaiveSeesaw}. While this may be possible with DUNE in the mass range below the kaon mass \cite{Krasnov:2019kdc,Ballett:2019bgd,Gunther:2023vmz} and SHiP \cite{Alekhin:2015byh,SHiP:2018xqw,Gorbunov:2020rjx} or FCC-ee \cite{FCC:2018evy,Blondel:2022qqo} can get very close, it is practically challenging, to say the least.\footnote{For masses below $\sim 1$ GeV, the HNL impact on neutrinoless double $\beta$ decay can provide a complementary probe \cite{Asaka:2011pb,Lopez-Pavon:2012yda,Drewes:2016lqo,Hernandez:2016kel,deVries:2024rfh}.}
This may rise fears that the perspective to constrain all eighteen parameters of the model with $n_s=3$ is hopeless. However, 
as the relations in appendix \ref{app:Uai2exp} reveal,
fortunately the larger dimensionality of the parameter space is at least partially alleviated by the fact that the next-to-leading term in the expansion of the $U_{\alpha i}^2$ in $\upepsilon$ can be of order $1/\sqrt{\upepsilon} = e^\gamma$ (and not of order one, as for $n_s=2$).

\begin{figure}
    \centering
    \includegraphics[width=.49\textwidth]{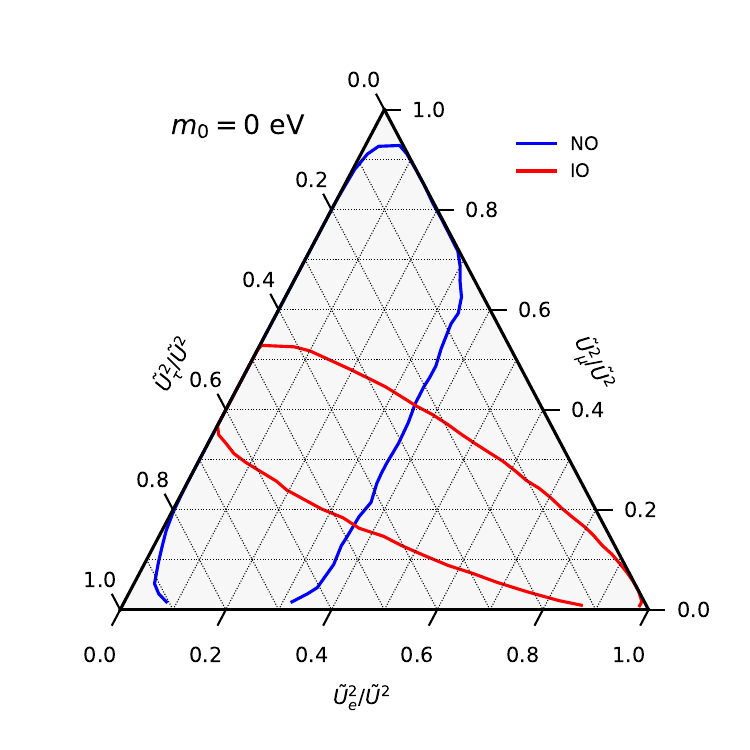}
    \caption{Range of flavour ratio $\tilde{U}_\alpha^2/\tilde{U}^2$ consistent neutrino oscillation data for $m_0 = 0$ eV when one allows for any values of the mixing angles in $V_\nu$ within $3\sigma$ \cite{Esteban:2020cvm}.}
    \label{fig:ternaryvanishingm0}
\end{figure}

The range of $\tilde{U}_\alpha^2/\tilde{U}^2$ that is consistent with light neutrino oscillation data is limited and depends on the mass of the lightest SM neutrino $m_0 = \operatorname{min}(m_1,m_3)$.
Hence, measuring both $m_0$ and the $\tilde{U}_\alpha^2/\tilde{U}^2$ provides a consistency check of the model. 
In Fig. \ref{fig:ternarygeneral_multiplem0} we display the allowed range of $\tilde{U}_\alpha^2/\tilde{U}^2$ for both orderings and different choices of $m_0$.
While similar plots have been shown in \cite{Chrzaszcz:2019inj,Krasnov:2023jlt} for $n_s=3$ we emphasise that using the parametrisation \eqref{eq:OROparametrisation} greatly reduces the computational effort needed to produce them (compared to the more common parametrisation of $\mathcal{R}$ in terms of three complex Euler angles) as the dependence on the two real angles $\theta_{N1},\theta_{N2}$ drops out. 
In Fig.~\ref{fig:ternaryvanishingm0} we display the range of $\tilde{U}_\alpha^2/\tilde{U}^2$ that is consistent with neutrino oscillation data for $m_0=0$, but allow the light neutrino mixing parameters to vary within the current $3\sigma$ intervals (in contrast to Fig.~\ref{fig:ternarygeneral_multiplem0} where we fixed them to their best fit value).

Considering the sums $\tilde{U}_\alpha^2$ does not only have the advantage that their ratios $\tilde{U}_\alpha^2/\tilde{U}^2$ are independent of several unknown parameters, but may also be practically necessary if the HNL mass splittings are too small to be resolved kinematically. In this case one may not be able to measure the $U_{\alpha i}$ independently, but constraints would apply to their sums $U_\alpha^2$.\footnote{Technically natural realisations of the model \eqref{L} with mixing angles that are large enough to be produced in sizeable numbers at the LHC (cf.~\textit{e.g.}~\cite{Drewes:2019fou}) tend to exhibit a quasi-degenerate mass spectrum \cite{Shaposhnikov:2006nn,Kersten:2007vk,Moffat:2017feq}. The largest mixing angles that lead to successful leptogenesis can be realised if all three HNL masses are quasi-degenerate \cite{Drewes:2021nqr}, and such a spectrum can also be motivated by discrete symmetries, see \textit{e.g.} \cite{Cirigliano:2005ck,Hagedorn:2006ug,Branco:2009by,Drewes:2022kap}.} 
In this work however we assume that the $M_i$ can be resolved individually.

In addition to $m_0$ the range of $\tilde{U}_\alpha^2/\tilde{U}^2$ can in principle also be constrained by restricting the range of $U^2$, cf.~Fig.~\ref{fig:Ua2triangle}. This is practically useful because $U^2$ (and hence $\tilde{U}^2$) can be obtained from the total number of events, and experiments probe large $U^2$ first and gradually slice through the $U^2$-axis as they keep taking data. However, the dependence of the flavour ratios on $U^2$ is relatively weak for NO and barely noticeable for IO which might make such test difficult in practice.

Though the assumption of arbitrary precision implies that contributions at any order in $\upepsilon$ can be resolved, we shall organise the following discussion in powers of $\upepsilon$.

\paragraph{Order $1/\upepsilon$.}

\begin{figure}[!t]
    \centering
    \begin{subfigure}[b]{0.5085\textwidth}
        \centering
        \includegraphics[width = \textwidth]{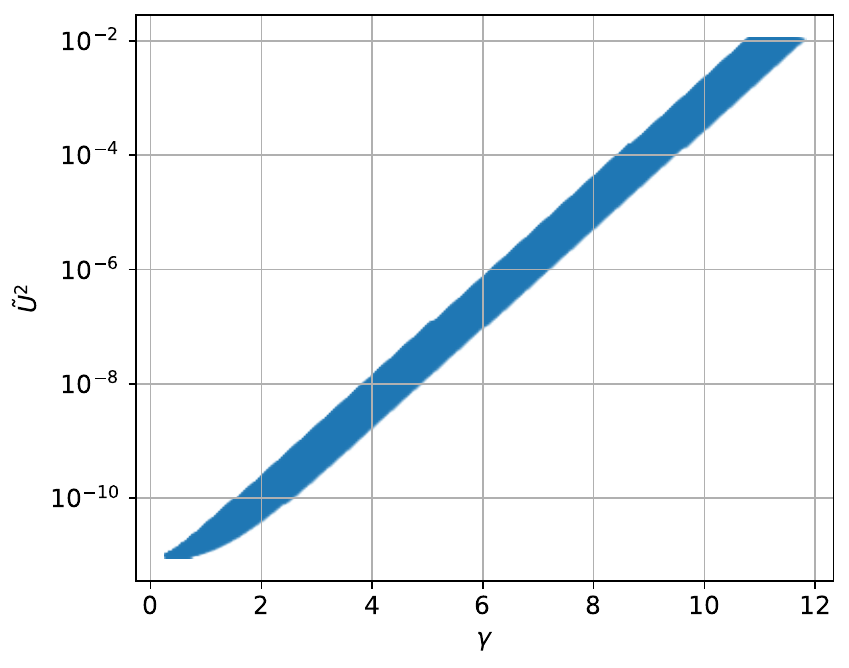}
        \subcaption{\label{fig:U2-gamma_range_NH}}
    \end{subfigure}
    \hfill
    \begin{subfigure}[b]{0.4815\textwidth}
        \centering
        \includegraphics[width = \textwidth]{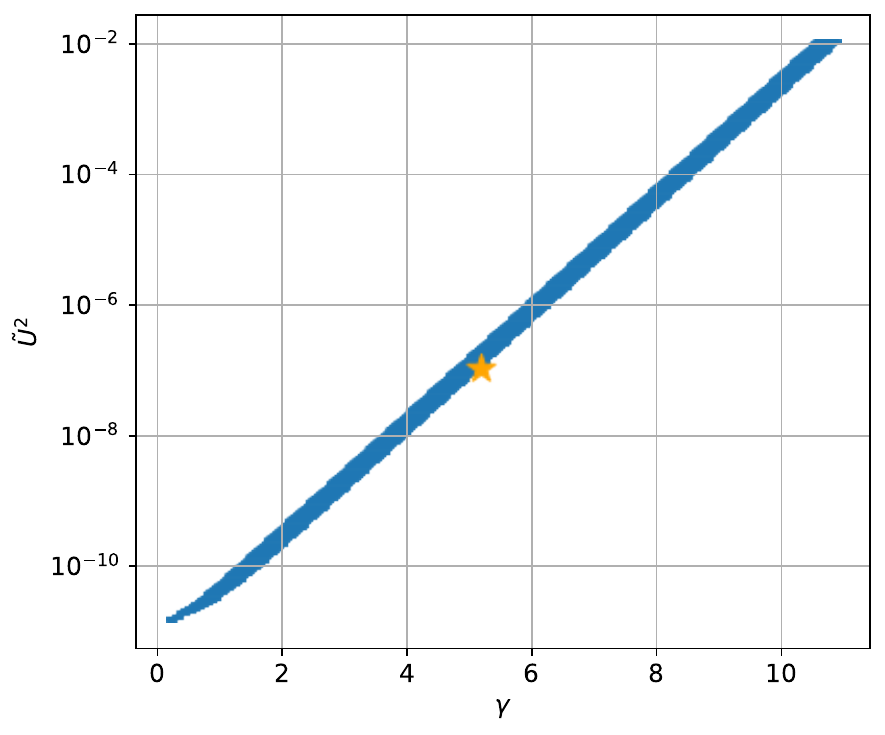}
        \subcaption{\label{fig:U2-gamma_range_IH}}
    \end{subfigure}
    \caption{\label{fig:U2-gamma_range}Allowed range for $\tilde{U}^2$ for each value of $\gamma$ using Eq.~\ref{eq:MU2} for normal ordering (left) and inverted ordering (right). Light neutrino oscillation data (including $\delta$) are fixed to their best fit values (with $m_0 = 0$ eV) while the rest of the model parameters are marginalised over. The orange star indicates the benchmark in Tab.~\ref{benchmark1}. 
    }
\end{figure}

\begin{figure}[!t]
    \centering
    \begin{subfigure}[b]{0.49\textwidth}
        \centering
        \includegraphics[width = \textwidth]{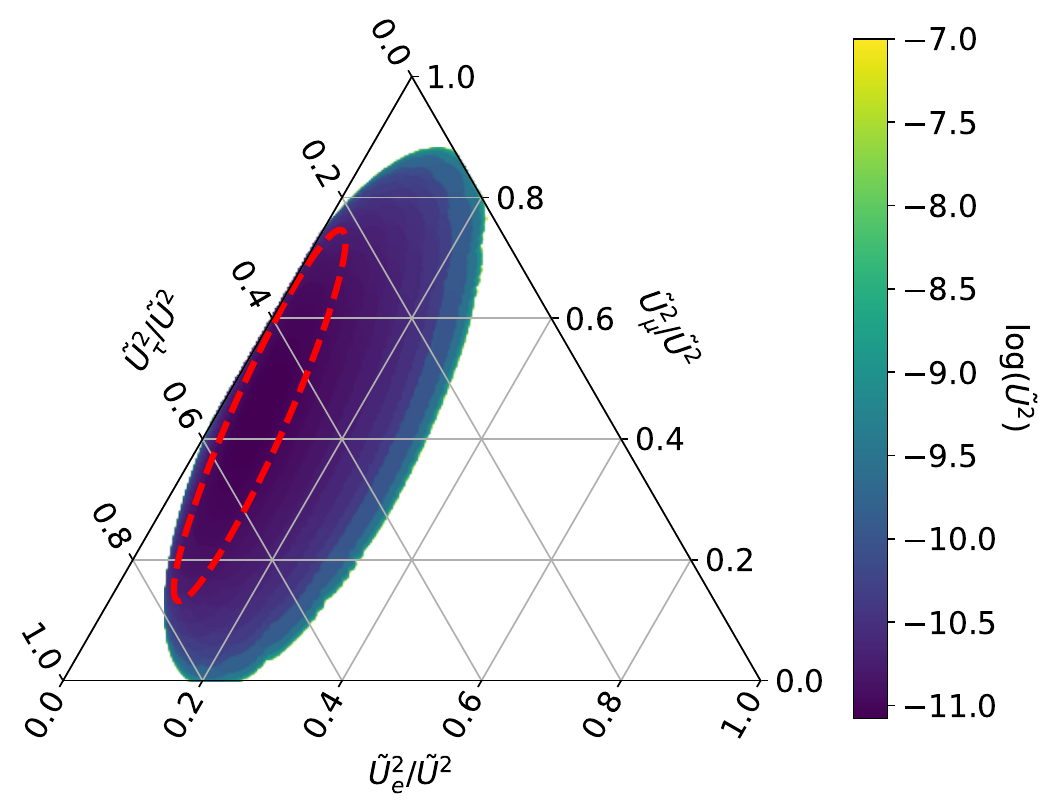}
        \subcaption{\label{fig:Ua2triangle_NH}}
    \end{subfigure}
    \hfill
    \begin{subfigure}[b]{0.49\textwidth}
        \centering
        \includegraphics[width = \textwidth]{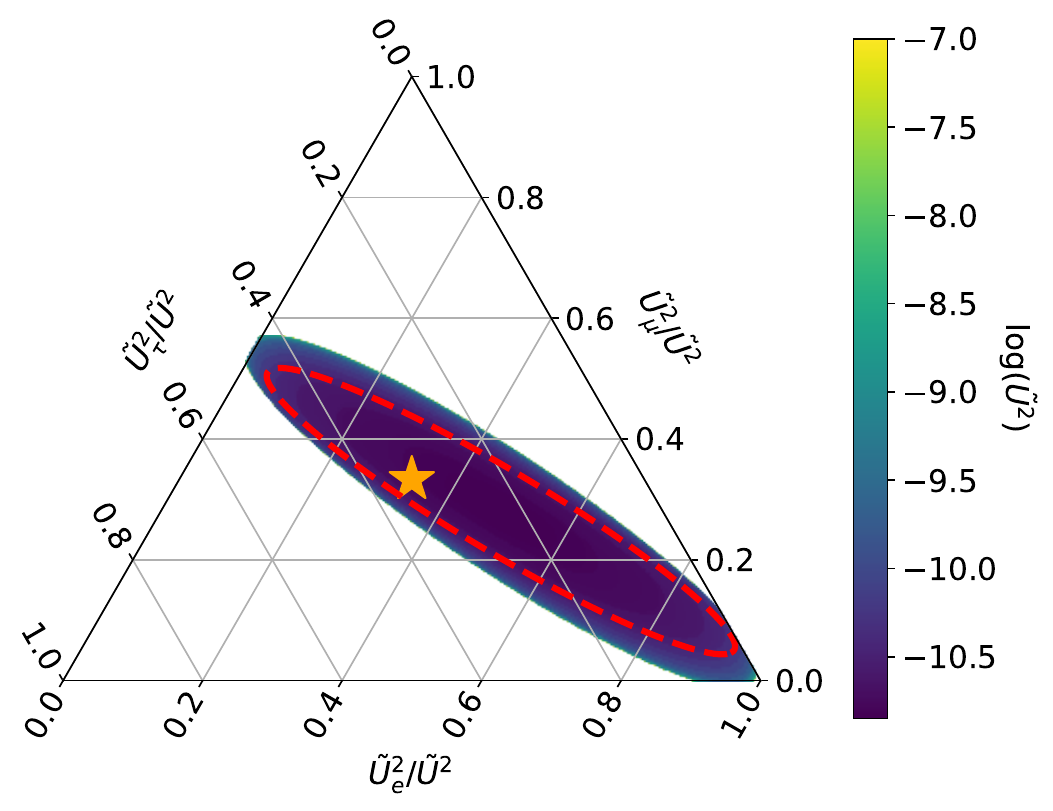}
        \subcaption{\label{fig:Ua2triangle_IH}}
    \end{subfigure}
    \caption{Range of allowed $\tilde{U}_{\alpha}^2 / \tilde{U}^2$ for $m_0=0$ and fixed light neutrino oscillation data (including $\delta$), obtained from \ref{eq:MUa2} for normal ordering (left) and inverted ordering (right). The orange star indicates the benchmark in Tab.~\ref{benchmark1}. The red dashed contour delimit the equivalent region in the $n_s=2$ scenario. The colourbar indicates the minimal value of $\tilde{U}^2$ that can be reached for specific flavour ratio. 
      \label{fig:Ua2triangle}}
\end{figure}

At leading order in the $\upepsilon$ expansion, the mixing angles factorise,
\begin{align}
	\label{eq:factorized}
	U^2_{\alpha i} = \frac{U_i^2 U_\alpha^2}{U^2} + \mathcal{O}(1/\sqrt{\upepsilon})\,.
\end{align}
Their sum has a very simple parametric dependence,
\begin{equation}
    \Mbar \tilde{\U}^2 
    = \Mbar \sum_{\alpha} \tilde{\U}_\alpha^2 = \frac{\Be^{2 \gamma}}{2} \big( \m_1[1-\cn^2\snn^2] + \m_2 \cn^2+\m_3[1-\cn^2 \cnn^2] \big)\label{eq:MU2}.
\end{equation}
In Figs.~\ref{fig:U2-gamma_range_NH} and \ref{fig:U2-gamma_range_IH} we show the relation between $\tilde{U}^2$ and $\gamma$. The width of the relation between these 2 quantities solely arise from the possible range of values taken by the $m_i$ and $\theta_{\nu i}$. 

\begin{figure}[!t]
    \centering
    \begin{subfigure}[b]{0.49\textwidth}
        \centering
        \includegraphics[width = \textwidth]{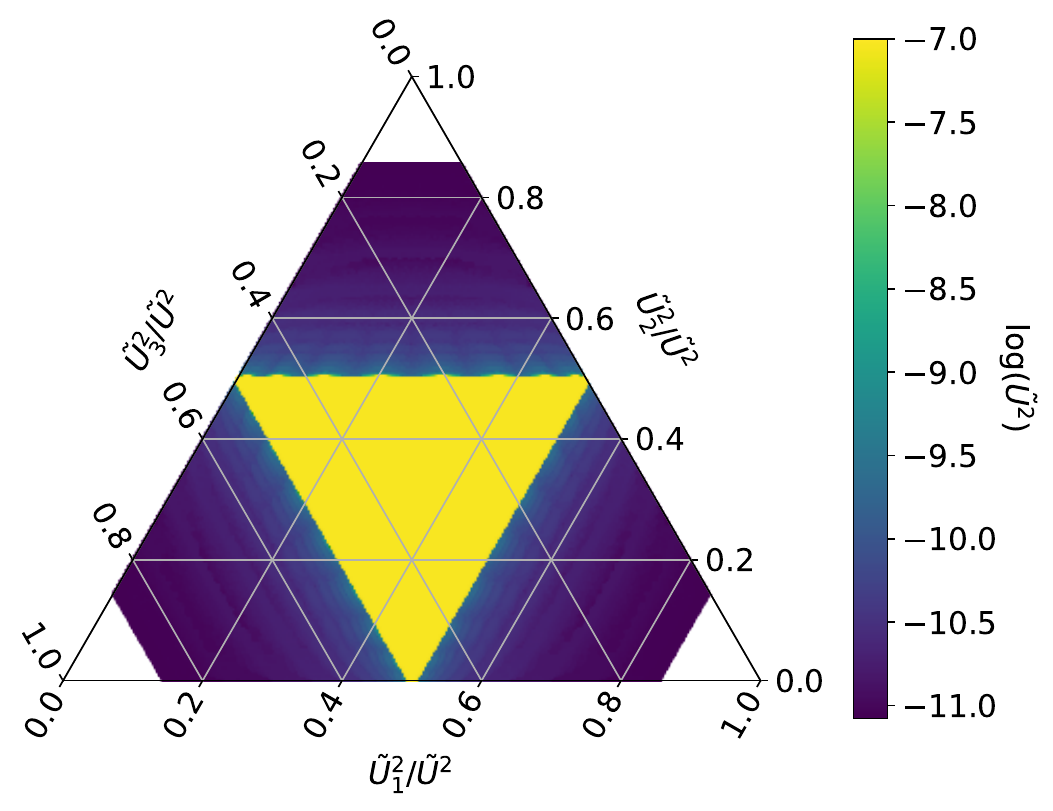}
        \subcaption{\label{fig:Ui2triangle_NH}}
    \end{subfigure}
    \hfill
    \begin{subfigure}[b]{0.49\textwidth}
        \centering
        \includegraphics[width = \textwidth]{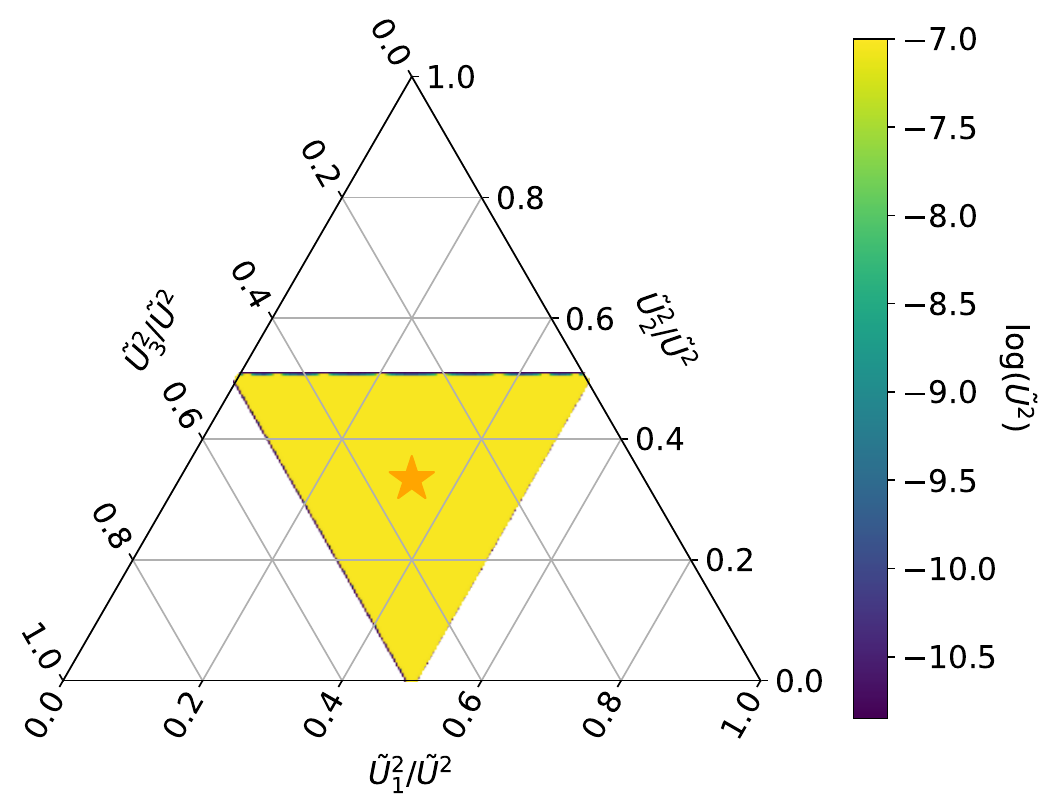}
        \subcaption{\label{fig:Ui2triangle_IH}}
    \end{subfigure}
    \caption{Range of allowed $\tilde{U}_i^2 / \tilde{U}^2$ for $m_0=0$ eV, obtained from Eq.~\eqref{eq:MiUis} for both normal ordering (left) and inverted ordering (right). The orange star indicates the benchmark presented in Tab.~\ref{benchmark1}. The colourbar indicates the maximal value of $\tilde{U}^2$ that can be reached for specific flavour ratios.
    \label{fig:Ui2triangle}}
\end{figure}

The individual terms $\tilde{\U}_\alpha^2$ read, at leading order in $\upepsilon$,
\begin{equation}
\begin{aligned}
    \Mbar \tilde{\U}_\alpha^2 &&&= \frac{\Be^{2 \gamma}}{2} \bigg( \m_1[1-\cn^2\snn^2] |\Vai{\alpha}{1}|^2 + \m_2 |\Vai{\alpha}{2}|^2 \cn^2+\m_3 |\Vai{\alpha}{3}|^2 [1-\cn^2 \cnn^2] \label{eq:MUa2}\\
    &&&+ 2\sqrt{\m_1 \m_3} \big[
    \sn \Im{\Vai{\alpha}{1} \Vai{\alpha}{3}^*} - \cn^2 \snn \cnn \Re{\Vai{\alpha}{1} \Vai{\alpha}{3}^*} \big] \\
    &&&+ 2\sqrt{\m_1 \m_2} \big[
    \cn \cnn \Im{\Vai{\alpha}{1} \Vai{\alpha}{2}^*} - \cn \sn \snn \Re{\Vai{\alpha}{1} \Vai{\alpha}{2}^*} \big] \\
    &&&- 2 \sqrt{\m_2 \m_3} \big[\cn \snn \Im{\Vai{\alpha}{2} \Vai{\alpha}{3}^*} + \cn \cnn \sn \Re{\Vai{\alpha}{2} \Vai{\alpha}{3}^*} \big] \bigg).
\end{aligned}
\end{equation}
In Fig.~\ref{fig:Ua2triangle}
we show the range of $\tilde{U}_\alpha^2/\tilde{U}^2$ that can be realised for given $U^2$.
Similarly, one can also restrict the range of $\tilde{U}_i^2/\tilde{U}^2$. At leading order, the $U_i^2$ are indeed related in the following way
\begin{equation}
  \tilde{\U}_i^2 =
  \tilde{\U}^2 \times
  \left\lbrace
  \begin{aligned}
      &\frac{1}{2} \big(1 - \cN^2 \sNN^2 \big)  & ~~~~  \mbox{for }i=1, \\
      &\frac{1}{2} \cN^2 & ~~~~ \mbox{for }i=2, \\
      &\frac{1}{2} \big(1 - \cN^2 \cNN^2 \big)  & ~~~~ \mbox{for }i=3.
  \end{aligned}
  \right.
  \label{eq:MiUis}
\end{equation}
The result is shown in Fig.~\ref{fig:Ui2triangle}.
Finally, we obtain
\begin{equation}\label{Eq3d8}
  M_i \Uai{\alpha}{i}^2 = \Mbar
  \tilde{\U}_{\alpha}^2 \times
  \left\lbrace
  \begin{aligned}
      &\frac{1}{2} \big(1 - \cN^2 \sNN^2 \big)  & ~~~~  \mbox{for }i=1, \\
      &\frac{1}{2} \cN^2 & ~~~~ \mbox{for }i=2, \\
      &\frac{1}{2} \big(1 - \cN^2 \cNN^2 \big) & ~~~~ \mbox{for }i=3,
  \end{aligned}
  \right.
\end{equation}
at leading order in $\upepsilon$. The above expressions show that the leading order term in $U_i^2$ could already give us information to potentially reconstruct five parameters

Indeed, out of the six quantities $U_i^2$ and $U_\alpha^2$ only five are independent, as the two sets both sum up to the same quantity $\tilde{U}^2$ in \eqref{eq:MU2Def}.
Hence, up to five parameters can be extracted from constraints on these combinations. 
This is possible analytically at order $1/\upepsilon$. 
To illustrate this, we consider the case $m_0=0$ with inverted ordering.
From \eqref{eq:MiUis} one can obtain $\theta_{\N 1}$ as well as $\theta_{\N 2}$, up to a discrete degeneracy.
Using \eqref{eq:MU2}, one can then express $\cn$ in terms of $\gamma$ and $\theta_{\nu 2}$.
Since there are two independent equations in \eqref{eq:MUa2}, one can use one of them to express $\alpha_2$ in terms of $\gamma$ and $\theta_{\nu 2}$, and substitute it into the other. 
Finally, when forming the combination $\tilde{U}_\alpha^2$, one can express $\theta_{\nu 2}$ in terms of $\gamma$.
Hence, we could express $\theta_{\N 1}$, $\theta_{\N 2}$, $\cn$, $\alpha_2$ and $\theta_{\nu 2}$ in terms of other parameters.
Due to the factorisation \eqref{eq:factorized} it is not possible to extract more information at order $1/\upepsilon$.

\paragraph{Order $1/\sqrt{\upepsilon}$.}
The appearance of $1/\sqrt{\upepsilon}$-enhanced contributions to the $U_{\alpha i}^2$ is a genuinely new feature for $n_s=3$ (compared to $n_s=2$, where only integer powers of $\upepsilon$ appear).
The leading $1/\upepsilon$-term  cancels when forming combinations such as $\tilde{\U}_{\alpha i}^2 \tilde{\U}_{\beta j}^2 - \tilde{\U}_{\beta i}^2 \tilde{\U}_{\alpha j}^2 $; hence, such combinations are sensitive to the NLO term. 
In order to extract the deviation from the form stated in~\eqref{eq:factorized}, it is instructive to construct combinations of observables that vanish in the factorised limit.
One such combination is given by the combinations
\begin{align}\label{NLOdistinctor}
	d_{\alpha i} = \sum_{j,k} \epsilon_{ijk} U_{\beta j}^2 U_{\gamma k}^2 = \Mbar^2\sum_{j,k} \epsilon_{ijk} \frac{\tilde{U}_{\beta j}^2 \tilde{U}_{\gamma k}^2}{M_jM_k}\,,
\end{align}
with $\alpha \neq \beta \neq \gamma$. Interestingly, the dependence on the mass splitting completely factorises in this expression.
The quantities \eqref{NLOdistinctor} scale distinctly differently for $n_s=2$ and $n_s=3$; 
\begin{align}
\label{eq:dalphai_ns=3}
	d_{\alpha i} \lesssim \mathcal{O}(U_0^4 e^{3 \gamma}) = \mathcal{O}(U^4 \sqrt{\upepsilon}) \quad {\rm with} \ n_s=3
\end{align}
 while we instead find for the analogue quantity\footnote{Strictly speaking, the definition \eqref{NLOdistinctor} does not make sense for $n_s = 2$ but one can define the equivalent quantity $d_{\alpha} = U_{\beta 1}^2 U_{\gamma 2}^2 - U_{\beta 2}^2 U_{\gamma 1}^2$, with $\alpha\neq \beta \neq \gamma$.} that
\begin{align}
\label{eq:dalphai_ns=2}
	d_{\alpha} = \mathcal{O}(U_0^4 e^{2 \gamma}) = \mathcal{O}(U^4 \upepsilon) \quad {\rm for} \ n_s=2.
\end{align}
Hence, they provide a smoking gun signature of a deviation from the minimal incarnation\footnote{If the measured $d_{\alpha i}$ largely differ from both the $n_s=2$ and $n_s=3$ estimates (see Eqs.~\eqref{eq:dalphai_ns=2} and \eqref{eq:dalphai_ns=3}), it could also hint towards additional new physics, \textit{e.g.} $n_s>3$ or interactions beyond the standard right-handed neutrino Yukawa coupling.} with $n_s=2$ even if the $\tilde{U}_\alpha^2/\tilde{U}^2$ happen to take values that would also be allowed in that scenario.

Quite remarkably, we find that all the model parameters are already accessible at $\mathcal{O}(1/\sqrt{\upepsilon})$ (instead of $\mathcal{O}(\upepsilon^0)$) for the benchmark that we consider in Tab.~\ref{benchmark1}.
This allows to compensate for the larger dimensionality of the $n_s=3$ scenario and is the reason why some of the real phases ($\theta_{\nu i},\omega$) can be better constrained compared to the $n_s=2$ case, see Sec.~\ref{sec:chi2method}. While we perform our analysis using the full system of equations, we emphasise that the $\mathcal{O}(1/\sqrt{\upepsilon})$ terms give the main constraints for these real phases.

\paragraph{The full picture.}
Our goal is to investigate whether there is a bijective mapping between the model parameters 
($\theta_{N1}$, $\theta_{N2}$, $\theta_{\nu 1}$, $\theta_{\nu 2}$, $\gamma$, $\omega$, $\alpha_1$ and $\alpha_2$)
and the $U_{\alpha i}^2$, assuming that the $M_i$ have been determined kinematically and the light neutrino masses $m_i$ as well as the mixing angles $\uptheta_{ij}$ and Dirac phase $\delta$ in \eqref{PMNS} have been determined in low energy experiments.  
When considering all orders in $\upepsilon$, the expressions for the $U_{\alpha i}^2$ in general become complicated and not particularly illuminating, with the exception of the combination
\begin{equation}
    \Mbar\tilde{\U}^2 = \cosh{2 \gamma} \big(\m_1[1-\cn^2\snn^2] +\m_2 \cn^2+\m_3[1-\cn^2 \cnn^2] \big) + \big ( \m_1\cn^2\snn^2 + \m_2\sn^2 + \m_3\cn^2\cnn^2 \big)
\end{equation}
Comparably simple expressions for other combinations of $U_{\alpha i}^2$ can be obtained in the limit $m_0\to 0$; they are given (at NLO, \textit{i.e.} at order $e^{\gamma} = 1/\sqrt{\upepsilon}$) in appendix \ref{app:Uai2exp}. In this limit the dimensionality of the parameter space is reduced, as one Majorana phase in \eqref{PMNS} becomes unphysical. 

The question whether all model parameters can be uniquely reconstructed from knowledge of the $U_{\alpha i}^2$ can only be investigated numerically. For this we define a first benchmark\footnote{
We truncate the numbers in Tab.~\ref{benchmark1} for convenience, the precise numerical values for the non-integer angles and phases $\theta_{\nu 1} = 0.22439948$,
$\theta_{\nu 2} = 0.97497703$,
$\theta_{N 1} = 0.61547971$,
$\theta_{N 2} = 0.78539816$,
$\alpha_2 = 0.31679926$, 
$\delta = 4.04916$.} 

\begin{table}[!h]
\centering
\begin{tabular}{ccccccccccc}
   $\Mbar$ &$\theta_{\nu 1}$ &$\theta_{\nu 2}$& $ \omega$ & $\gamma$ & $\theta_{\N 1}$&$\theta_{N2}$ & $\delta$  & $\alpha_1$&$\alpha_2$
    &$ m_0$
    \\ 
    \hline
 $7$ GeV & $\pi/14$ &$\pi/3.2$& $ 1$ & $5$ & $\pi/5.1$ & $\pi/4$ & $1.3\pi$  & $0$ & $\pi/10$
    &$0$ eV
    \end{tabular}\nonumber\\
\caption{\label{benchmark1}
Parameter values for the benchmark indicated in Figs.~\ref{fig:Ua2triangle} and \ref{fig:Ui2triangle} with an inverted ordering of the light neutrino masses.
}
\end{table}

Fixing all other parameters to their best fit value in NuFit 5.2 \cite{Esteban:2020cvm}, the mixing angles $U_{\alpha i}^2$ take, in this scenario, the following values

\begin{align}
\label{Uaimatrix1}
    \tilde{U}_{\alpha i}^2=\begin{pmatrix}
    1.15053679 & 1.11098616 & 1.14804674 \\
    1.11216418 & 1.12135875 & 1.11711550 \\
    1.12395834 & 1.13097212 & 1.13695779 \\    
    \end{pmatrix} \cdot 10^{-8}
\end{align}

corresponding to a position approximately in the middle of both triangles displayed in Figs.~\ref{fig:Ua2triangle} and \ref{fig:Ui2triangle}, consistent with one of the benchmark configurations proposed in \cite{Drewes:2022akb}. 
It is also useful to define a second benchmark with non-zero $m_0$,

\begin{table}[!h]
\centering
\begin{tabular}{ccccccccccccc}
    $\Mbar$ &$\theta_{\nu 1}$ &$\theta_{\nu 2}$& $ \omega$ & $\gamma$ & $\theta_{\N 1}$&$\theta_{N2}$ & $\delta$  & $\alpha_1$&$\alpha_2$
    &$ m_0$
    \\ 
    \hline
  $7$ GeV & $\pi/2.57$ &$\pi/3.2$& $ 1$ & $5$ & $\pi/5.1$ & $\pi/4$ & $1.3\pi$  & $0$ & $\pi/10$
    &$0.01$ eV
    \end{tabular}\nonumber\\
\caption{Parameter values for the benchmark with $m_0\neq 0$. Inverted ordering is assumed. \label{benchmark2}}
\end{table}

leading to

\begin{align}
\label{Uaimatrix2}
    \tilde{U}_{\alpha i}^2=\begin{pmatrix}
        1.02268430 & 1.05553109 & 1.03596221 \\
    0.297159580 & 0.280769175 & 0.282038087 \\
    1.91813382 & 1.88980908 & 1.91967347 \\
    \end{pmatrix}\cdot 10^{-8}.
\end{align}

We then perform a numerical scan of the model parameter space to see if 
the parameter values in
Tabs.~\ref{benchmark1} and \ref{benchmark2} are the unique choices (within the fundamental domain defined in Sec.~\ref{sec:modelandparam})  that lead to the matrices in Eqs.~\eqref{Uaimatrix1} and \eqref{Uaimatrix2}, respectively.
We find that this is indeed the case, \textit{i.e.}, the mapping between the model parameter values and the values of the $U_{\alpha i}^2$ is bijective.\footnote{We verified that such a mapping also exists for benchmark points with normal ordering.}
This implies that a determination of all $U_{\alpha i}^2$ (\textit{e.g.}~from flavoured branching ratios in HNL decays) would in principle be sufficient in order to determine all parameters in the Lagrangian \eqref{L}, provided that the light neutrino masses $m_i$, mixings $\uptheta_{ij}$ and the Dirac phase $\delta$ are determined in low energy experiments. 
Hence, the answer to question \ref{it:inPrinciple} posed in the introduction is positive. 

However, two important caveats need attention. Firstly, the result was obtained numerically for two specific choices of true parameters.
We checked that it is correct for a few more choices, but it cannot be concluded that the bijectivity holds everywhere in the model's parameter space. 
Secondly, uniquely identifying the parameter values in Tab.~\ref{benchmark1} from the $\tilde{U}_{\alpha i}^2$ in Eq.~\eqref{Uaimatrix1} requires using the full expressions given in appendix \ref{app:Uai2exp}. If one only relies on the leading order terms in $\upepsilon$, then there are discrete degeneracies. For instance, $\theta_{N 1} \to \pi - \theta_{N 1}$ is an approximate symmetry of the system. 
Since difference arising from subleading terms in $\upepsilon$ are difficult to detect experimentally, it raises the question how much information can be obtained with realistic experimental sensitivities.

\section{What can  experiments do semi-realistically?}\label{sec:chi2method}

Given the conclusions from the previous section, a natural question to ask is whether near futures experiments can realistically measure the $U_{\alpha i}^2$ with the sufficient precision. 
So far we have phrased the discussion in terms of the $\tilde{U}_{\alpha i}$ and combinations thereof, in particular the ratios \eqref{Uratios}, which are independent of the HNL masses $M_i$. In order to compute actual event numbers in a given experiment, the $M_i$ must be fixed, and it is common to define benchmarks in terms of the ratios $U_\alpha^2/U^2$ instead (cf.~\textit{e.g.}~\cite{Drewes:2018gkc,SHiP:2018xqw,Tastet:2021vwp,Beacham:2019nyx,Drewes:2022akb}).
For the proof of principle that we aim for we can assume that we are in a regime where the differences between the $M_i$ are large enough to make them kinematically distinguishable\footnote{In the following, we will also neglect any uncertainty coming from the finite resolution of these mass splittings and assume they are known exactly.} (implying that the $U_{\alpha i}^2$ can be measured individually), but small enough not to significantly modify neutrino oscillation data, in particular justifying to approximate $\tilde{U}_{\alpha i}^2/\tilde{U}^2\simeq U_{\alpha i}^2/U^2$ in what follows. 
Obviously this approximation would not be justified in an exhaustive parameter scan.

Conceptually the ability of accelerator-based experiments to constrain the $U_{\alpha i}^2$ is limited by two factors. Firstly, the total number of HNL events is limited by the number of collisions, introducing a  statistical uncertainty. Secondly, the ability to extract information from a given number of event is limited by the properties of the detector in use.
Addressing both of these issues requires detailed simulations not only of the collisions, but also of the detectors, which goes beyond the scope of the present work. 
We restrict ourselves to an estimate of the purely statistical uncertainties due to the finite number of events, \textit{i.e.}, we address question \ref{it:EventNumbers}, and leave question \ref{it:Detectors} for experiment-specific studies in the future. 

Based on the expansion in $\upepsilon$, the most promising experiments are those that can probe mixing angles close to the seesaw-line \eqref{NaiveSeesaw} in the mass-mixing plane. 
Values of the $M_i$ exceeding a few tens of MeV are constrained by a combination of direct searches \cite{Abdullahi:2022jlv}
and cosmological 
constraints from primordial nucleosynthesis
and the subsequent history \cite{Sabti:2020yrt,Boyarsky:2020dzc,Vincent:2014rja,Diamanti:2013bia,Poulin:2016anj,Domcke:2020ety}. 
The three proposed experiments that can get closest to \eqref{NaiveSeesaw}
are then searches in the DUNE near detector \cite{Krasnov:2019kdc,Ballett:2019bgd}, the SHiP experiment \cite{SHiP:2018xqw,Gorbunov:2020rjx} and long-lived particle searches at FCC-ee or CEPC \cite{Blondel:2014bra,Blondel:2022qqo,CEPCStudyGroup:2018ghi}.
For our analysis we pick FCC-ee/CEPC because the number of events that can be seen in an idealised detector of given dimensions can be computed analytically \cite{Drewes:2022rsk}, and because formulae to estimate the sensitivities to the $U_{\alpha i}^2$ readily exist \cite{Antusch:2017pkq}. 
Note, however, that SHiP would e.g.~have a good sensitivity to probe the HNL properties \cite{Tastet:2019nqj,Mikulenko:2023olf}.

\subsection{Statistical method}

Our evaluation of statistical uncertainties in experiments is inspired by the approach taken in \cite{Antusch:2017pkq} and assumes background-free searches with event numbers given by the analytical estimates in \cite{Drewes:2022rsk}, which is a reasonable first approximation in displaced vertex searches at lepton colliders. 
We assume that all charged particles in all final states can be seen with $100\%$ detector efficiencies. In practice the extraction of the $U_{\alpha i}^2$ is most straightforward from semi-leptonic final states from charged current mediated HNL decays, as there the flavour of the charged lepton always corresponds to $\alpha$.

\paragraph{Approximate $\chi^2$-distributions.}
 We shall use the following estimate of the $\chi^2$-function to assess the statistical significance of measurements 
\begin{equation}
    \chi^2 = \sum_{\alpha, i} \left[\frac{(\U_{\alpha i}^2)_{measured} - (\U_{\alpha i}^2)_{evaluated}}{\Delta(\U_{\alpha i}^2)_{measured}}\right]^2, 
\end{equation}
where $(\U_{\alpha i}^2)_{measured}$ indicates a value of $U_{\alpha i}^2$ extracted from measurements with a statistical uncertainty $\Delta(\U_{\alpha i}^2)_{measured}$, and $(\U_{\alpha i}^2)_{evaluated}$ is the fiducial value, \textit{i.e.}, the true value of $U_{\alpha i}^2$ for a specified set of benchmark parameter values. 
Let  ${\rm N}_{\alpha i}$ be the number of measured events mediated by the mixing $U_{\alpha i}^2$, \textit{i.e.} arising from the decay of the HNL $N_i$ into any lepton of flavour $\alpha$. We also define ${\Nsl} = \sum_{\alpha, i} {\rm N}_{\alpha i}$ the total number of observed semi-leptonic events.
Given that the number of events in one specific channel ${\rm N}_{\alpha i}$ behaves as a Poisson process \cite{Antusch:2017pkq}, the statistical uncertainties on the measured quantities can be estimated as 
\begin{equation}
    \frac{\Delta(\U_{\alpha i}^2)}{\U_{\alpha i}^2} \simeq \frac{\sqrt{\text{N}_{\alpha i}}}{\text{N}_{\alpha i}} \ , \quad 
    \frac{\U_{\alpha i}^2}{\U^2} \simeq \frac{\text{N}_{\alpha i}}{\Nsl}  \quad
    \implies \Delta(\text{U}_{\alpha i}^2) = \sqrt{\frac{\text{U}_{\alpha i}^2 \text{U}^2 }{\Nsl}} \ , \quad
\end{equation}
leading to
\begin{equation}
    \chi^2 = \Nsl\sum_{\alpha, i} \frac{\left[(\text{U}_{\alpha i}^2)_{measured} - (\text{U}_{\alpha i}^2)_{evaluated}\right]^2}{(\text{U}_{\alpha i}^2 \text{U}^2)_{measured}}.
    \label{eq:chi2func}
\end{equation}
For the benchmarks in Tabs.~\ref{benchmark1} and \ref{benchmark2} we find, based on the equations in \cite{Drewes:2022rsk} with an IDEA-type detector \cite{FCC:2018evy} and with the idealised assumption that all final states can be reconstructed,
that $\Nsl\simeq 10^5$ events should be observed. We use this number in the following. With dedicated long-lived particle detectors \cite{Wang:2019xvx,Chrzaszcz:2020emg} an even larger number could potentially be produced.

\paragraph{Sensitivity to NLO terms.}

\begin{figure}[!t]
  \centering
  \includegraphics[angle=0,width=0.49\textwidth]{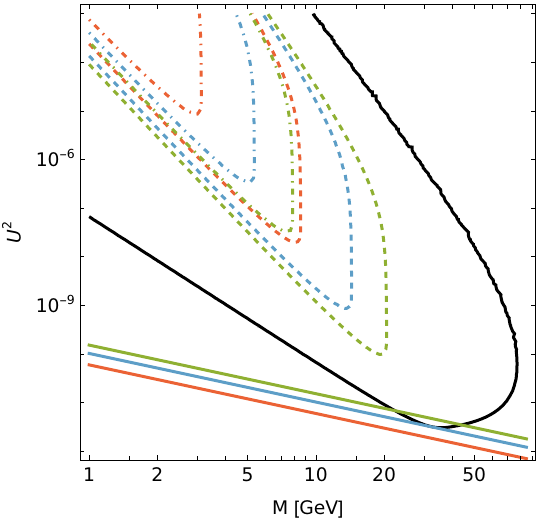}
	\caption{Expected sensitivity to measuring the parameters $d_{\alpha i}$ based on the estimates for $n_s=3$ in~\eqref{eq:sensitivity_estimate_N3}.
		The solid black line corresponds to the expected reach of FCC-ee or CEPC \cite{Blondel:2022qqo,Drewes:2022rsk}. The coloured lines correspond to the minimal seesaw lines \eqref{NaiveSeesaw} 
  for $M_1=M_2=M_3=M$, on one hand
  with $m_0=0$ for normal ordering (red) or inverted ordering (blue), and on the other hand for the maximal sum of the light neutrino masses $\sum m_\nu$ allowed by cosmological bounds \cite{Planck:2018vyg} (green). The dashed (dot-dashed) lines correspond to measuring \eqref{NLOdistinctor} at the $3\sigma$ ($5 \sigma$) level. 
	}
  \label{fig:sensEst}
\end{figure}

The absence of terms $\propto 1/\sqrt{\upepsilon}$ is a distinctive feature of the minimal realistic incarnation of \eqref{L} with $n_s=2$, where only integer powers of $\upepsilon$  appear in the expansion of the $U_{\alpha i}^2$. They do appear for $n_s=3$, and it is instructive to pose the question where in the HNL mass-mixing plane FCC-ee or CEPC can be sensitive to these terms. 
We use the quantities \eqref{NLOdistinctor} to estimate this sensitivity.
Assuming that the variables $U_{\alpha i}^2$ are independent, we find that
\begin{align}
	\frac{\delta d_{\alpha i}}{d_{\alpha i}} = \frac{1}{\sqrt{\Nsl}} \frac{\mathcal{O}(U^4)}{d_{\alpha i}}\,,
\end{align}
or specifically for the case with $n_s=3$ we find the scaling 
\begin{align}
	\label{eq:sensitivity_estimate_N3}
	\frac{\delta d_{\alpha i}}{d_{\alpha i}} \lesssim \frac{1}{\sqrt{\mathrm{N}}} \mathcal{O}\left(\sqrt{\frac{U^2}{U^2_0} }\right)\,.
\end{align}
This is in stark contrast to the case with $n_s=2$ where the signal scales as
\begin{align}
	\frac{\delta d_{\alpha}}{d_{\alpha}} = \frac{1}{\sqrt{\mathrm{N}}} \mathcal{O}\left(\frac{U^2}{U^2_0} \right)\,.
\end{align}
We illustrate the region where FCC-ee or CEPC can be expected to be sensitive at the $3\sigma$ and $5\sigma$ level to the $1/\sqrt{\upepsilon}$-terms in Fig.~\ref{fig:sensEst}. For SHiP or the DUNE near detector, an analyic estimate of the expected event numbers cannot be performed in the same way as for FCC-ee or CEPC, and
a detailed simulation of the detector would be needed to make a similar analysis.\footnote{Semi-analytic estimates of the event numbers also exist for dump experiments \cite{Bondarenko:2019yob} and for the LHC \cite{Drewes:2019vjy}, but are more complicated and less accurate. }
 For SHiP, the expected number of up to $10^4$ events above the kaon mass (see figure 10 in \cite{Tastet:2019nqj}) along with the fact that the sensitivity is governed by the distance to the seesaw line, see Eq.~\eqref{eq:dalphai_ns=3}, indicates a similar sensitivity.
The DUNE experiment could reach $\mathcal{O}(10)$ events even for $U^2\sim U_0^2$, for which such NLO terms are no longer suppressed.

\subsection{FCC-ee or CEPC sensitivities to individual model parameters}

In the following we use  \eqref{eq:chi2func} to estimate the error bars with which various model parameters for the benchmark in Tab.~\ref{benchmark1} can be determined in long-lived particle searches at FCC-ee or CEPC under the aforementioned assumptions. 
For the one-dimensional $\chi^2$-functions we indicate the regions with $\chi^2 = 1, 4, 9$ by dashed lines; 
the region where the blue curve remains below those lines roughly corresponds to the $1\sigma$, $2\sigma$ and $3\sigma$ intervals for the respective parameters. 

\begin{figure}[!t]
    \centering
    \includegraphics[width = 0.49\textwidth]{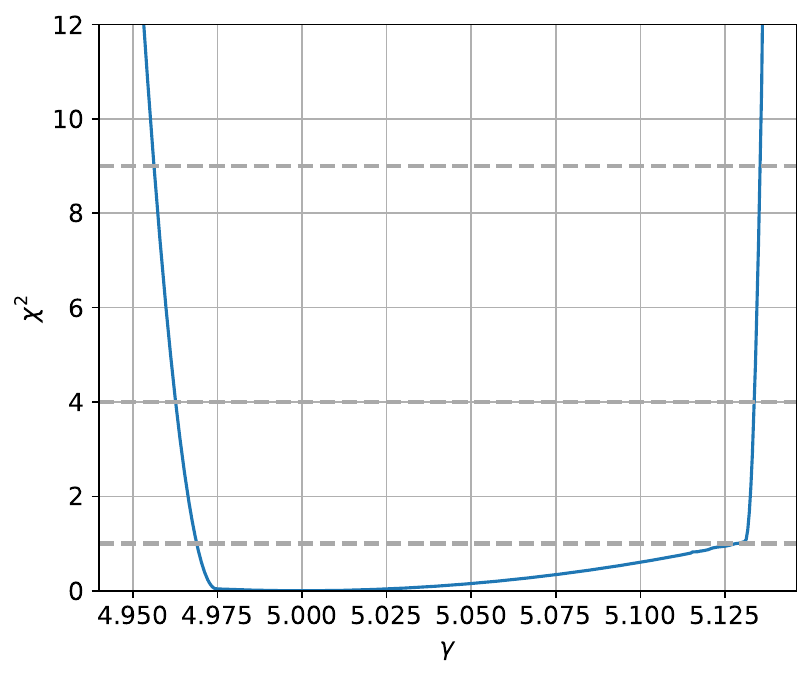}
    \includegraphics[width = 0.49\textwidth]{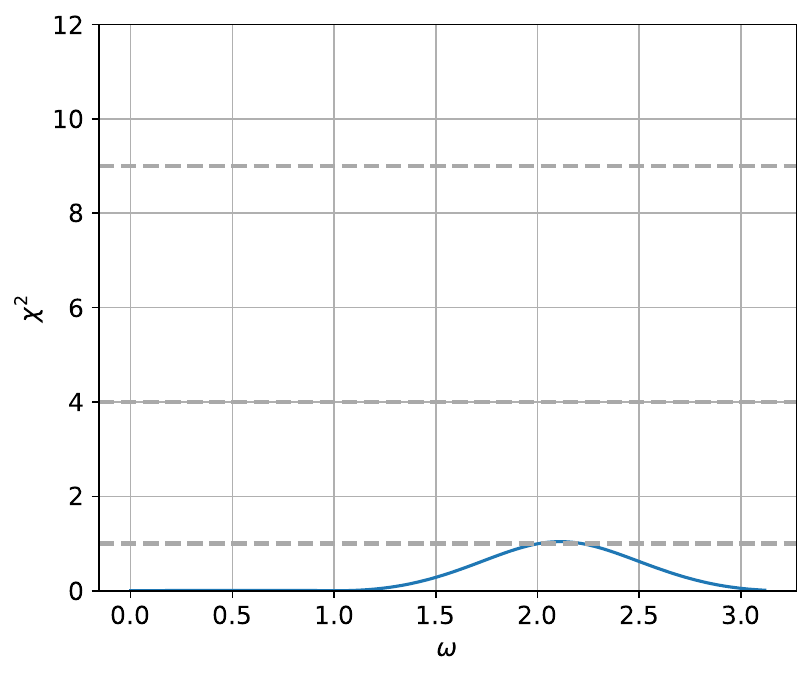}
    \caption{(\textit{Left panel}) Dependence of the $\chi^2$-function on $\gamma$. All other model parameters are marginalised over. (\textit{Right panel}) Equivalent plot for $\omega$.
    }
    \label{fig:gamma_and_omega}
\end{figure}

The left panel in Fig.~\ref{fig:gamma_and_omega} shows that the imaginary part $\gamma$ of the angle in $\mathrm{R_C}$ can be extracted with a very good precision, which one may have expected from \eqref{eq:MU2} and Fig.~\ref{fig:U2-gamma_range}.
The real part $\omega$, on the other hand, remains largely unconstrained, cf. the right panel of Fig.~\ref{fig:gamma_and_omega}.
This comes expected, as the analogue to $\omega$ is known to be the parameter that is most difficult to measure for $n_s=2$; the fact that at least some range can be disfavoured at the $1\sigma$ level is, in comparison, good news. For lower HNL masses (in the range of DUNE or SHiP) one may be able to obtain information on $\omega$ from neutrinoless double beta decay (cf.~\cite{deVries:2024rfh} and references therein for an updated analysis), but for the masses in our benchmark this does not seem realistic. 

\begin{figure}[!t]
    \centering
    \includegraphics[width = 0.502\textwidth]{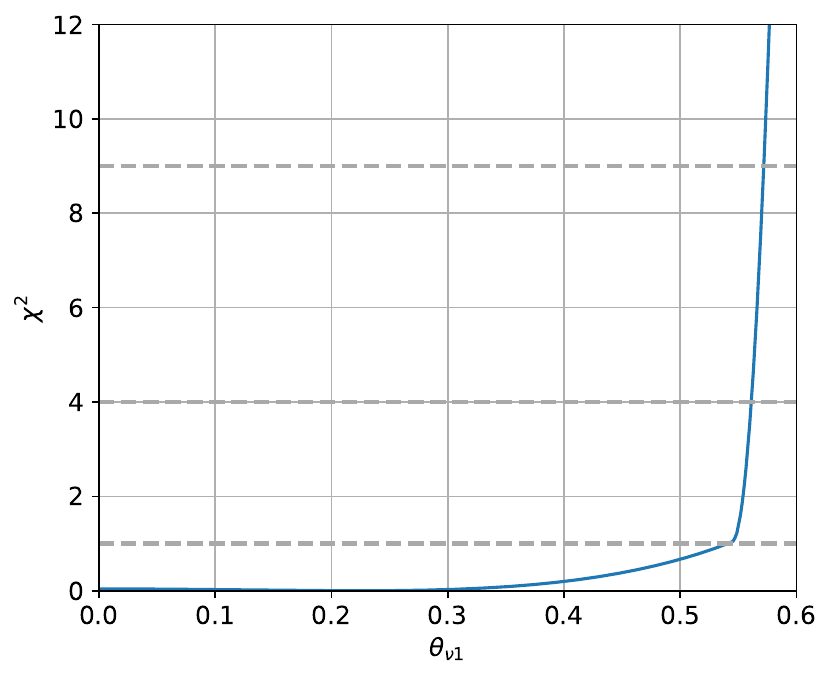}
    \includegraphics[width = 0.49\textwidth]{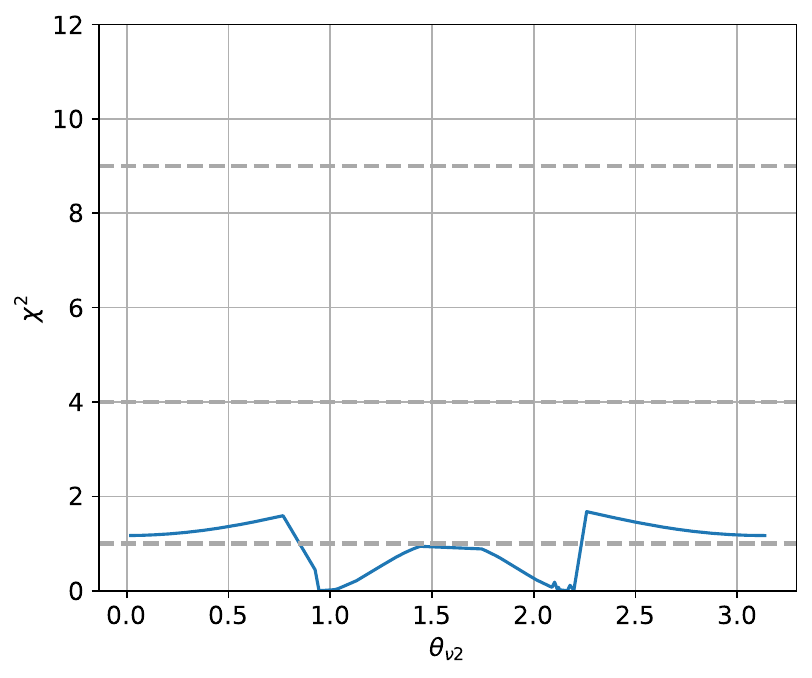}
    \caption{(\textit{Left panel}) Dependence of the $\chi^2$-function on $\theta_{\nu 1}$. All other model parameters are marginalised over. (\textit{Right panel}) Equivalent plot for $\theta_{\nu 2}$. 
    }
    \label{fig:tn1_and_tn2}
\end{figure}

\begin{figure}[!t]
    \centering
    \includegraphics[width = 0.49\textwidth]{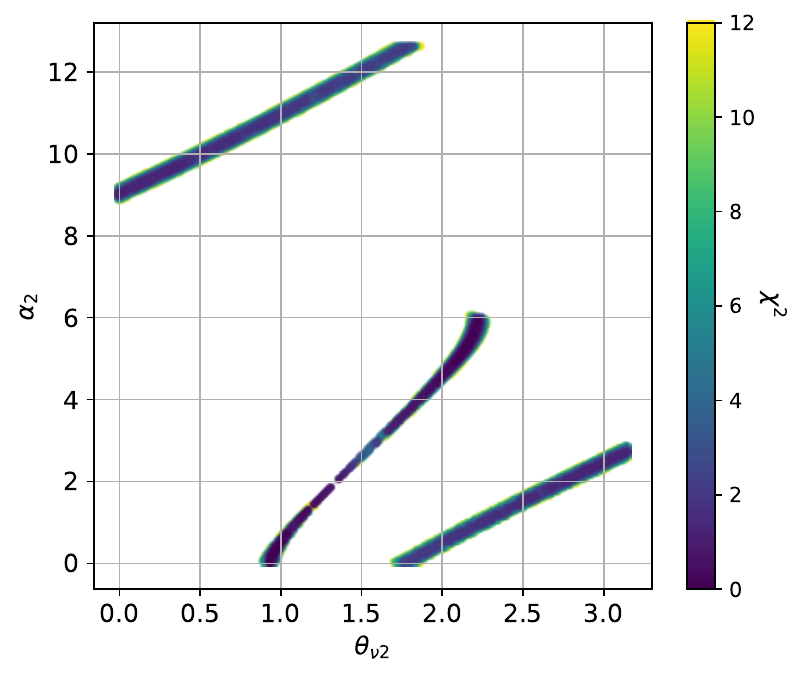}
     \includegraphics[width = 0.49\textwidth]{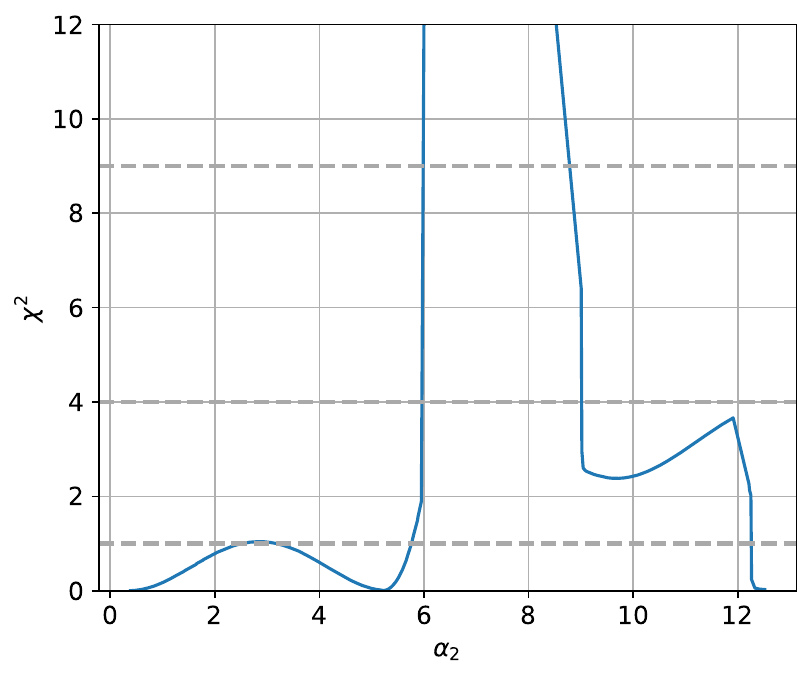}
    \caption{(\textit{Left panel}) Two-dimensional $\chi^2(\theta_{\nu 2}, \alpha_2)$, represented by the colours as indicated in the legend. (\textit{Right panel}) Dependence of the $\chi^2$-function on $\alpha_2$. In both figures, all remaining model parameters are marginalised over.}
    \label{fig:a2}
\end{figure}

The situation for the real angles in $\O_\nu$ is only slightly better, as can be seen in Fig.~\ref{fig:tn1_and_tn2}.
It should, however, be pointed out that the one-dimensional $\chi^2$ in this case is of limited use, as considerable correlations with the Majorana phase $\alpha_2$ can be seen in the left panel of Fig.~\ref{fig:a2}. 
A marginalised one-dimensional $\chi^2$ for $\alpha_2$ is shown in the right panel of Fig.~\ref{fig:a2}. The fact that some regions can be excluded is not surprising, as the possibility to constrain the Majorana phase from measurements of the $U_{\alpha i}^2$ is already known from the model with $n_s=2$ \cite{Drewes:2016jae,Caputo:2016ojx}; for $n_s=3$ this is somewhat hampered by the correlation visible in the left panel of Fig.~\ref{fig:a2}.

\begin{figure}[!t]
    \centering
    \includegraphics[width = .9\textwidth]{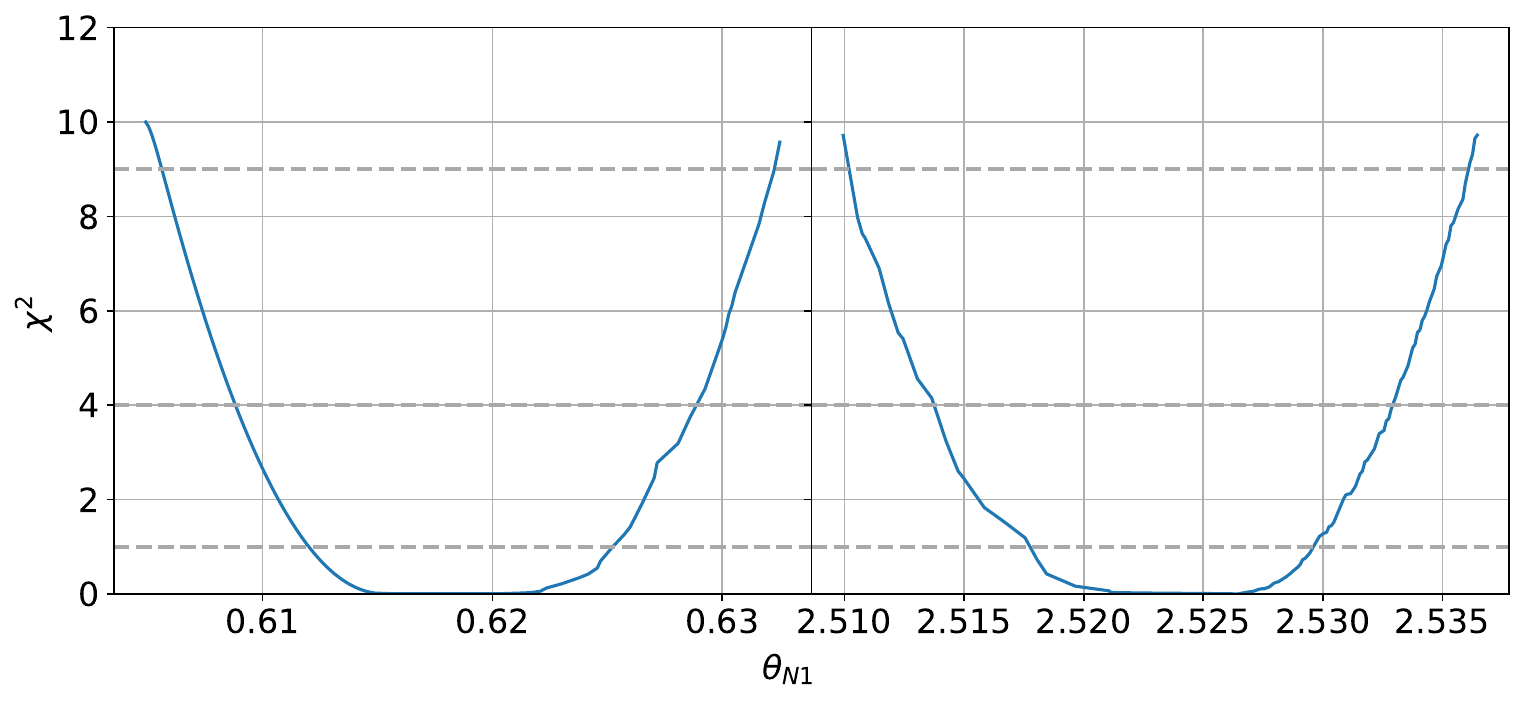}
    \caption{Dependence of the $\chi^2$-function on $\theta_{N1}$. All other model parameters are marginalised over.
    }
    \label{fig:tN1}
\end{figure}

\begin{figure}[!t]
    \centering
    \includegraphics[width = 0.49\textwidth]{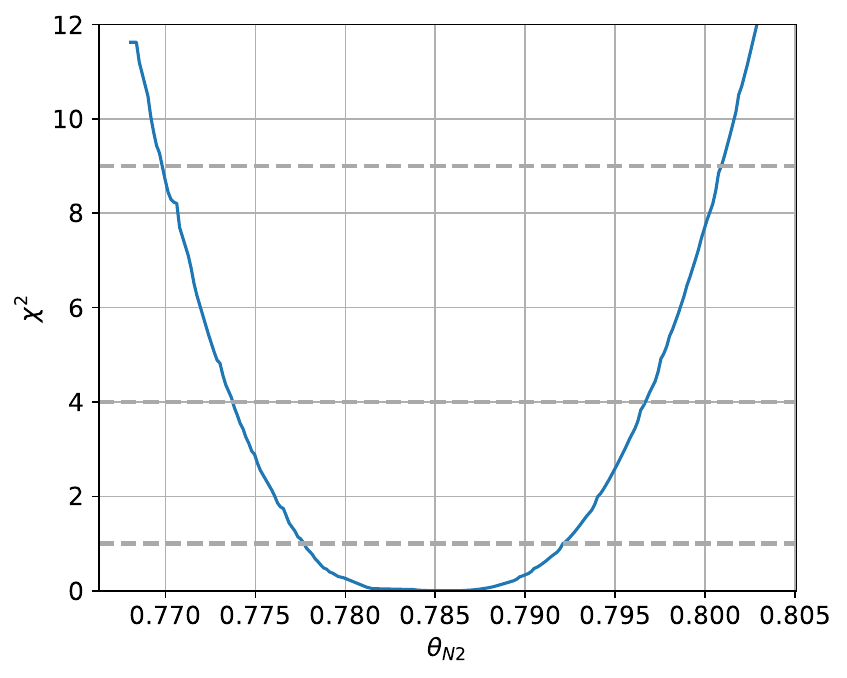}
    \caption{Dependence of the $\chi^2$-function on $\theta_{N2}$. All other model parameters are marginalised over.
    }
    \label{fig:tN2}
\end{figure}

Figs.~\ref{fig:tN1} and \ref{fig:tN2} show that
the angles in $\O_\N$ can, on the other hand, can be constrained very well. There are, however, two minima the lower one of which cannot be identified with the $10^5$ events assumed here. This is a result of the aforementioned approximate symmetry $\theta_{N 1} \to \pi - \theta_{N 1}$ at leading order in $\upepsilon$, which is only broken by higher order terms.

Deviating from our benchmark, it could also be interesting to look at the constraints that the $U_{\alpha i}^2$-measurements can set on the value of the Dirac CP-phase $\delta$ if the latter is not considered as fixed to its best fit value. Unfortunately, we observed it is only possible to put some (weak) constraints at the $1\sigma$-level on $\delta$, weaker than the constraints that are expected from DUNE and T2HyperK in the near-future.

\subsection{The lightest neutrino mass}

The benchmark in Tab.~\ref{benchmark1} has the convenient property $m_0=0$ eV, which eliminated one Majorana phase in \eqref{PMNS}. While this was sufficient as a proof-of-principle that experiments can constrain many of the model parameters, the question of how the situation changes with $m_0>0$ is inevitable. 
In general, letting $m_0$ vary does not qualitatively modify the conclusions regarding experimental sensitivities to other parameters.

\begin{figure}[!t]
    \centering
\includegraphics[width=.49\textwidth]{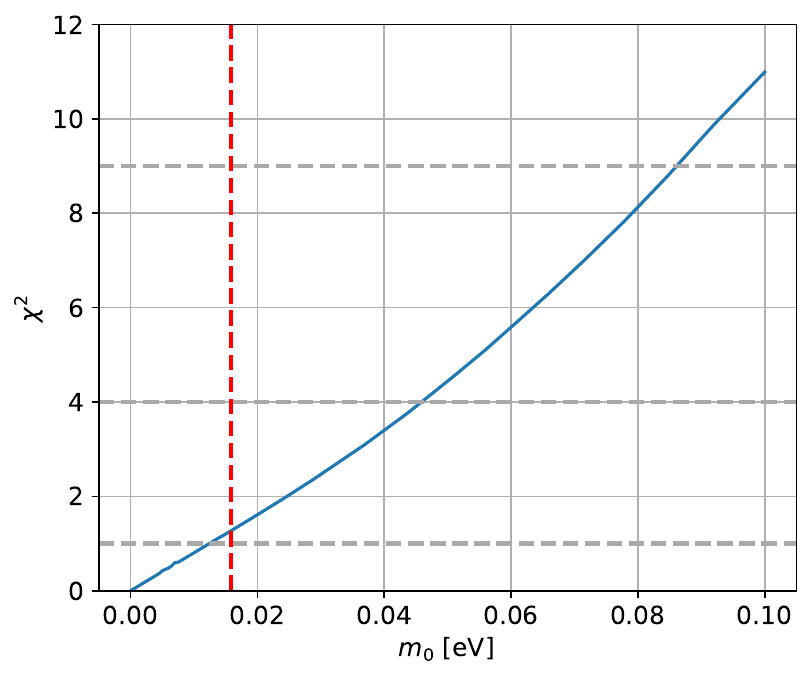}
\includegraphics[width=.49\textwidth]{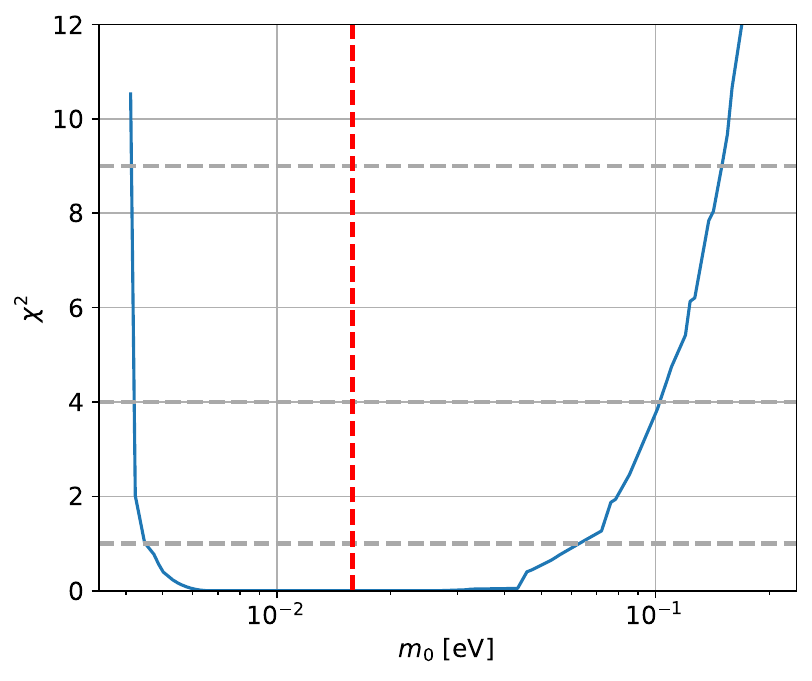}
    \caption{Dependence of the $\chi^2$-function on the lightest neutrino mass $m_0$ for an inverted ordering of the light neutrino masses. The true value of $m_0$ is chosen to be $0$ eV ($0.01$ eV) for the left (right) panel. We marginalised over the rest of the model parameters. The red dashed line denotes the constraints from Planck on this parameter \cite{Planck:2018vyg}.}
    \label{fig:m0_constraints}
\end{figure}

We consider the case where $m_0$ is not determined by other experiments, but has to be reconstructed from collider data.
Such an analysis is displayed in Fig.~\ref{fig:m0_constraints}. 
In the left panel, we use the benchmark defined in Tab.~\ref{benchmark1}, assuming $m_0 = 0$ eV as the true value, and 
quantify the upper bound on the 
lightest neutrino mass that can be obtained from measurements of the $U_{\alpha i}^2$ at future lepton colliders (assuming $10^5$ observed events). 
For comparison, the constraint currently set by Planck on $\sum_i m_i$ \cite{Planck:2018vyg} translates into  $m_0\leq 0.015$ eV (for inverted ordering). 
We observe that $m_0>0.015$ eV can only be excluded for this benchmark at the $1\sigma$ level. 
It is therefore unlikely that colliders will be competitive with cosmological constraints. 
It is still however very desirable to have complementary laboratory constraints on the value of $m_0$. In that regard, the current best constraint is set by KATRIN by looking at the tritium $\beta$ decay spectrum, limiting $m_0 \lesssim 0.45$ eV at 90\% CL \cite{Katrin:2024tvg}.\footnote{Strictly speaking the quantity constrained by KATRIN is not $m_0$, but the so-called electron neutrino mass $\sqrt{\sum_i m_i^2 |(V_\nu)_{ai}|^2}$. However, in the regime of the current sensitivity of the experiment, this quantity represents a good proxy for $m_0$.} 
Hence, the constraints on $m_0$ set by colliders have the potential to improve by an order of magnitude the current bound. Future improvements of the KATRIN constraints are expected to reach $0.2$ eV \cite{KATRIN:2005fny}.
Hence, future lepton colliders can at least in principle be more sensitive to $m_0$ than KATRIN. However, it should of course be kept in mind that such a measurement is indirect and model dependent, while the KATRIN constraints (being directly based on kinematics) are much less model-dependent. Moreover, in contrast to the sensitivity estimates of KATRIN, our analysis does not account for realistic detector simulations.

A similar analysis can be done for our second benchmark, summarised in Tab.~\ref{tab:param_restriction2}, which is similar in all respects to our first benchmark except for 1) the lightest neutrino mass which is set to $m_0 = 0.01$ eV 2) the angles $\theta_{\nu 1}$ and $\theta_{\nu 2}$ which are chosen in a way that the position of this benchmark in the ternary plot, shown in Fig.~\ref{fig:ternarygeneral_multiplem0}, lies outside of the $m_0 = 0$ eV-region. This enhances our prospect of discriminating between $m_0 = 0$ and $m_0\neq 0$. The dependence of the $\chi^2$-function on $m_0$ is shown in the right panel of Fig.~\ref{fig:m0_constraints}. Interestingly, we observe that it is possible to exclude the massless limit $m_0 = 0$ at more than $3\sigma$. Beyond the framework of the type-I seesaw with 3 right-handed neutrinos, such a measurement at FCC-ee/CEPC would either provide a lower bound on the lightest neutrino mass $m_0$ or hint towards the presence of additional physics responsible for cancellations in the light neutrino mass matrix.

\section{Discussion and conclusion}
\label{sec:conclusion}

Heavy right-handed neutrinos appear in many popular extensions of the SM and can potentially resolve several open puzzles in particle physics and cosmology, including the light neutrino oscillation, the baryon asymmetry of the universe and the Dark Matter puzzle. 
From an experimental viewpoint, the presence of right-handed neutrinos would manifest itself as a type of HNL $N_i$ that mixes with the SM neutrinos. 
If the $N_i$ are lighter than the weak gauge bosons, then they can be produced in large numbers at colliders even if their sole interaction with ordinary matter is through the fractional SM weak interaction given to them by this mixing. 

Amongst the most promising way to find those HNLs are displaced vertex searches, where the expected number of events can in good approximation be expressed in terms of the $U_{\alpha i}^2$ defined in \eqref{UaiDefinition}, which are commonly used to define benchmark scenarios. If any HNLs are discovered in experiments, a key question will be to understand their role in particle physics and cosmology, and in particular the connection to models of neutrino mass. In the popular type-I seesaw model this connection can readily be expressed in terms of the $U_{\alpha i}^2$ or, more conveniently, in terms of their combinations $\tilde{U}_\alpha^2$ and $\tilde{U}_i^2$ defined in \eqref{UiTildeDef} and \eqref{eq:MU2Def}. 
In the present work we addressed the question of how much about the connection to an underlying model of neutrino mass generation can be learnt from measuring the HNL masses $M_i$ and mixings $U_{\alpha i}^2$. Working in the seesaw model \eqref{L} with three HNL flavours ($n_s=3$), we generalised previous studies in the minimal model with two HNL flavours ($n_s=2$). 

As a first step in Sec.~\ref{sec:inprinciple_learning} we identified correlations between the allowed range of the ratios $\tilde{U}_\alpha^2/\tilde{U}^2$ and the lightest neutrino mass $m_0$ (Figs.~\ref{fig:ternarygeneral_multiplem0} and \ref{fig:ternaryvanishingm0}) or the total mixing $\tilde{U}^2$ (Fig.~\ref{fig:Ua2triangle}). 
While similar plots have previously been shown in the literature, we emphasise that the realisation \eqref{eq:OROparametrisation} of the Casas-Ibarra parametrisation \eqref{CAparam} is particularly useful in this context, as it does not only make explicit that the deviation of the $U_{\alpha i}^2$ from the so-called seesaw line \eqref{NaiveSeesaw} is governed by one single parameter $1/\upepsilon$ \eqref{UpepsilonDef}, but also is suitable for generating such plots with little numerical effort.  
We also found correlations for the ratios
$\tilde{U}_i^2/\tilde{U}^2$ and $\tilde{U}^2$ in Fig.~\ref{fig:Ui2triangle}.
These correlations provide a powerful test of the neutrino mass model.

We then proceeded to address the question of how many of the new model parameters can be determined uniquely from the $M_i$ and $U_{\alpha i}^2$, provided that the light neutrino masses and mixing angles as well as the Dirac phase have been determined in low energy experiments. We find that, if the $U_{\alpha i}^2$ could be measured with arbitrary precision, then all model parameters could be uniquely reconstructed in the benchmark summarised in Tab.~\ref{benchmark1}. 
Expanding the $U_{\alpha i}^2$ in powers of the parameter $|\upepsilon| < 1$, we find that five model parameters can already be determined at order $1/\upepsilon$.
A key difference to the minimal model with $n_s=2$ is that information about the remaining parameters becomes available already at order $1/\sqrt{\upepsilon}$. This permits using the combinations \eqref{NLOdistinctor} as a smoking gun to detect a deviation from the minimal model even if the $\tilde{U}_\alpha^2/\tilde{U}^2$ happen to lie in a regime that would also be allowed for $n_s=2$, cf.~fig.~\ref{fig:sensEst}. We caution that the existence of a bijective mapping between model parameters and the mixings $U_{\alpha i}^2$ was only verified for a few benchmark points without an exhaustive scan of the parameter space. It is clear that there exists lower-dimensional submanifolds of the parameter space where the parameter reconstruction is not as advantageous,  the prime example being the subspace where one effectively recovers the $n_s=2$ scenario, for which such a bijective mapping does not exist. However, we observed that its existence is a generic feature of all the benchmark points that we considered, which were not purposefully chosen to support this conclusion.

In Sec,~\ref{sec:chi2method} we then estimated how many model parameters can be reconstructed from experimental data with a finite number of events in the benchmark in Tab.~\ref{benchmark1}, assuming idealised detectors and considering statistical uncertainties only. We considered the case of FCC-ee or CEPC, which can potentially produce over $10^5$ HNLs in displaced vertex searches at the $Z$-pole within the parameter space allowed by current bounds. 
Fig.~\ref{fig:gamma_and_omega} shows that the imaginary part $\gamma$ of the angle in $\mathrm{R_C}$ can be measured rather precisely, while constraining the real part $\omega$ is very difficult. 
Constraining the angles in $\O_\nu$ is also challenging (Fig.~\ref{fig:tn1_and_tn2}), while those in $\O_\N$ can be measured (Figs.~\ref{fig:tN1} and \ref{fig:tN2}), and for the Majorana phase $\alpha_2$ at least a sizeable region can be excluded. It should, however, also be kept in mind that there are correlations between these parameters, cf.~Fig.~\ref{fig:a2}.
We also discuss the observation that, while $\delta$ can also in principle be extracted from the measurements of these mixing angles, the limits set by said measurements are in practice weak. 
Finally, in Fig.~\ref{fig:m0_constraints} we considered a different benchmark and found that the FCC-ee or CEPC could potentially measure the lightest neutrino mass $m_0$ with a precision comparable to current cosmological bounds and better than KATRIN, though this measurement would of course be indirect and more model-dependent that at KATRIN.

While the potential of a given experiment to discover HNLs is typically understood in terms of its sensitivity to smaller $U_{\alpha i}^2$, the ability to distinguish different models by studying the branching ratios in their decays also crucially depends on $\upepsilon$.
Since $\upepsilon$ is parametrically related to the distance to the so-called seesaw line, cf.~\eqref{eq:enhancement}, this distance can be used as an indicator for an experiment's ability to test the underlying model in case HNLs are discovered. 
The most promising existing experiments in this regard are NA62 and the LHC main detectors. The proposed and planned facilities that can minimise the distance to the seesaw line are, in addition FCC-ee/CEPC discussed here, the DUNE near detector and SHiP.

In summary, we used the example of FCC-ee or CEPC to show that next-generation accelerator-based experiments do not only have the potential to discover HNLs, but can also discriminate the minimal and next-to-minimal incarnations of the seesaw models from each other in case of a discovery. They can further measure many of the new model parameters, providing an important step forward to understand the role of HNLs in particle physics and cosmology.

\acknowledgments

YG acknowledges the support of the French Community of Belgium through the FRIA grant No.~1.E.063.22F and thanks the UNSW School of Physics for its hospitality during part of this project.

\begin{appendix}

\section{Analytical expressions \label{app:Uai2exp}}

In this appendix, we provide the analytical expressions for the mixing angles $U_{\alpha i}^2$ (or linear combinations of these) which we used to extract model parameters in the analysis described in Sec.~\ref{sec:chi2method}. For the sake of brevity, we focus on the case where the lightest neutrino is massless ($m_0 = 0$ eV).

For both normal and inverted ordering, we will need the following 8 quantities
\begin{equation}
    \begin{aligned}
        \r1{\pm} &= \cNN\snn[\sn^2\pm\cn^2]-\sN\sNN\sn\cnn, \\
        \r2{\pm} &= \cNN\cnn[\sn^2\pm\cn^2]+\sN\sNN\sn\snn, \\
        \r3{\pm} &= \cNN\cnn\sn + \sN\sNN\snn[\sn^2\pm\cn^2], \\
        \r4{\pm} &= \cNN\snn\sn - \sN\sNN\cnn[\sn^2\pm\cn^2] .
    \end{aligned}
\end{equation}

\subsection{Normal ordering}

Starting with normal ordering, the $\Uai{\alpha}{i}^2$ can simply be written, in the case of $\m_1 = 0$ eV, as

\begin{align}
        \Mbar \tilde{\U}^2 = \sum_{\alpha, i} M_i \Uai{\alpha}{i}^2 =&~ \frac{\Be^{2 \gamma}}{2} \big( \m_2 \cn^2+\m_3[1-\cn^2 \cnn^2] \big)\\
 \Mbar \tilde{\Ua{\alpha}}^2 = \sum_{i} M_i \Uai{\alpha}{i}^2 =&~ \frac{\Be^{2 \gamma}}{2} \bigg( \m_2 |\Vai{\alpha}{2}|^2 \cn^2+\m_3 |\Vai{\alpha}{3}|^2 [1-\cn^2 \cnn^2] \\
    &- 2 \sqrt{\m_2 \m_3} \big[\cn \snn \Im{\Vai{\alpha}{2} \Vai{\alpha}{3}^*} + \cn \cnn \sn \Re{\Vai{\alpha}{2} \Vai{\alpha}{3}^*} \big] \bigg) \notag\\
    M_2 \Ui{2}^2 =&~ \Mbar \tilde{\U}^2 \left(\frac{\cN^2}{2} \right) \label{eq:MU22NO} \\
    &+ \Be^\gamma \cN \sN \bigg( \cos{\omega} \big[- \m_2 \cn \sn + \m_3 \cn \sn \cnn^2 \big] + \sin{\omega} \m_3 \cn \cnn \snn \bigg) \notag\\
    M_2 \Uai{\alpha}{2}^2 =&~ \Mbar \tilde{\Ua{\alpha}}^2 \left(\frac{\cN^2}{2} \right) \label{eq:MUa22NO} \\
    &+ \Be^\gamma \cN \sN \bigg(\cos{\omega} \bigg[- \m_2 |\Vai{\alpha}{2}|^2 \cn \sn + \m_3 |\Vai{\alpha}{3}|^2 \cn \sn \cnn^2  \notag\\
    &+ \sqrt{\m_2 \m_3} ( \cnn[\sn^2-\cn^2] \Re{\Vai{\alpha}{2} \Vai{\alpha}{3}^*} + \sn \snn \Im{\Vai{\alpha}{2} \Vai{\alpha}{3}^*} )\bigg] \notag\\
    &+ \sin{\omega} \bigg[\m_3 |\Vai{\alpha}{3}|^2 \cn \cnn \snn \notag\\
    &+ \sqrt{\m_2 \m_3} (\sn \snn \Re{\Vai{\alpha}{2} \Vai{\alpha}{3}^*} - \cnn \Im{\Vai{\alpha}{2} \Vai{\alpha}{3}^*} ) \bigg] \bigg) \notag \\
    M_1 \Ui{1}^2 =&~ \Mbar \tilde{\U}^2 \left(\frac{1 - \cN^2\sNN^2}{2} \right) \label{eq:MU12NO}\\
    &+ \Be^\gamma \cN \sNN \bigg(\cos{\omega} \big[\m_2  \cn \sN \sNN \sn + \m_3 \cn \cnn \r1+ \big] \notag\\
    &+ \sin{\omega} \big[\m_2  \cn \cNN \sn - \m_3  \cn \cnn \r3+ \big]\bigg)\notag\\
    M_1 \Uai{\alpha}{1}^2 =&~ \Mbar \tilde{\Ua{\alpha}}^2 \left(\frac{1 - \cN^2\sNN^2}{2} \right) \label{eq:MUa12NO} \\
    &+ \Be^\gamma \cN \sNN \bigg(\cos{\omega} \bigg[\m_2 |\Vai{\alpha}{2}|^2 \cn \sN \sNN \sn + \m_3 |\Vai{\alpha}{3}|^2 \cn \cnn \r1+ \notag\\
    &+ \sqrt{\m_2 \m_3}(\r4- \Re{\Vai{\alpha}{2} \Vai{\alpha}{3}^*} - \r2+ \Im{\Vai{\alpha}{2} \Vai{\alpha}{3}^*}) \bigg] \notag\\
    &+ \sin{\omega} \bigg[\m_2 |\Vai{\alpha}{2}|^2 \cn \cNN \sn - \m_3 |\Vai{\alpha}{3}|^2 \cn \cnn \r3+ \notag\\
    &+ \sqrt{\m_2 \m_3}(-\r2- \Re{\Vai{\alpha}{2} \Vai{\alpha}{3}^*} - \r4+ \Im{\Vai{\alpha}{2} \Vai{\alpha}{3}^*}) \bigg] \bigg).\notag
\end{align}

\subsection{Inverted ordering}

Equivalently, for inverted ordering, the $\Uai{\alpha}{i}^2$ write, in the case of $\m_3 = 0$ eV, as

\begin{align}
    \Mbar \tilde{\U}^2 = \sum_{\alpha, i} M_i \Uai{\alpha}{i}^2 =&~ \frac{\Be^{2 \gamma}}{2} \big( \m_1 (1-\cn^2\snn^2) + \m_2 \cn^2 \big)\\
    \Mbar \tilde{\Ua{\alpha}}^2 = \sum_{i} M_i \Uai{\alpha}{i}^2 =&~ \frac{\Be^{2 \gamma}}{2} \bigg( \m_1 |\Vai \alpha1|^2 (1-\cn^2\snn^2) + \m_2 |\Vai{\alpha}{2}|^2 \cn^2 \\
    &+ 2 \sqrt{\m_1 \m_2} \big[\cn \cnn \Im{\Vai{\alpha}{1} \Vai{\alpha}{2}^*} - \cn \sn \snn \Re{\Vai{\alpha}{1} \Vai{\alpha}{2}^*} \big] \bigg) \notag\\
    M_2 \Ui{2}^2 =&~ \Mbar \tilde{\U}^2 \left(\frac{\cN^2}{2} \right) \label{eq:MU22IO} \\
    &+ \Be^\gamma \cN \sN \bigg(\cos{\omega} \bigg[\m_1  \cn\sn\snn^2 - \m_2  \cn\sn \bigg] \notag\\
    &+ \sin{\omega} \bigg[- \m_1 \cn \cnn \snn \bigg] \bigg) \notag\\
    M_2 \Uai{\alpha}{2}^2 =&~ \Mbar \tilde{\Ua{\alpha}}^2 \left(\frac{\cN^2}{2} \right) \label{eq:MUa22IO} \\
    &+ \Be^\gamma \cN \sN \bigg(\cos{\omega} \bigg[\m_1 |\Vai \alpha1|^2 \cn\sn\snn^2 - \m_2 |\Vai \alpha2|^2 \cn\sn \notag\\
    &+ \sqrt{\m_1 \m_2} ( [\sn^2-\cn^2]\snn \Re{\Vai{\alpha}{1} \Vai{\alpha}{2}^*}
    + \sn \cnn \Im{\Vai{\alpha}{1} \Vai{\alpha}{2}^*} )\bigg] \notag\\
    &+ \sin{\omega} \bigg[- \m_1 |\Vai \alpha1|^2\cn \cnn \snn \notag\\
    &+ \sqrt{\m_1 \m_2} (- \sn \cnn \Re{\Vai{\alpha}{1} \Vai{\alpha}{2}^*}
    +\snn \Im{\Vai{\alpha}{1} \Vai{\alpha}{2}^*}
     ) \bigg] \bigg) \notag\\
    M_1 \Ui{1}^2 =&~ \Mbar \tilde{\U}^2 \left(\frac{1 - \cN^2\sNN^2}{2} \right) \label{eq:MU12IO}\\
    &+ \Be^\gamma \cN \sNN \bigg(\cos{\omega} \bigg[\m_2 \sN \sNN \cn \sn  -\m_1 \cn \snn \r2+ \bigg]\notag\\
    &+ \sin{\omega} \bigg[\m_2  \cNN\cn\sn  -  \m_1 \r4+ \bigg] \bigg)\notag\\
    M_1 \Uai{\alpha}{1}^2 =&~ \Mbar \tilde{\Ua{\alpha}}^2 \left(\frac{1 - \cN^2\sNN^2}{2} \right) \label{eq:MUa12IO} \\
    &+ \Be^\gamma \cN \sNN \bigg(\cos{\omega} \bigg[\m_2|\Vai \alpha2|^2 \sN \sNN \cn \sn  -\m_1|\Vai \alpha1|^2 \cn \snn \r2+ \notag\\
    &+ \sqrt{\m_1 \m_2}(-\r3- \Re{\Vai{\alpha}{1} \Vai{\alpha}{2}^*}
    +\r1+ \Im{\Vai{\alpha}{1} \Vai{\alpha}{2}^*}) \bigg] \notag\\
    &+ \sin{\omega} \bigg[\m_2 |\Vai \alpha2|^2 \cNN\cn\sn  -  \m_1 |\Vai \alpha1|^2\r4+  \notag\\
    &+ \sqrt{\m_1 \m_2}(-\r1- \Re{\Vai{\alpha}{1} \Vai{\alpha}{2}^*}
    - \r3+ \Im{\Vai{\alpha}{1} \Vai{\alpha}{2}^*}) \bigg] \bigg)\notag
\end{align}

\end{appendix}

 \bibliographystyle{JHEP}
\bibliography{biblio}

\end{document}